%% file: TOP-11-015_temp.tex
\pdfoutput=1

\documentclass[11pt,twoside,a4paper,cmspaper,final,collab]{cms-tdr}
\def\svnVersion{177020}\def\cmsCernNo{2012-250}\def\cmsCernDate{\today}\def\cmsMessage{Submitted to the Journal of High Energy Physics}

\begin{document}\cmsNoteHeader{TOP-11-015}

\hyphenation{had-ron-i-za-tion}
\hyphenation{cal-or-i-me-ter}
\hyphenation{de-vices}

\RCS$Revision: 156083 $
\RCS$HeadURL: svn+ssh://svn.cern.ch/reps/tdr2/papers/TOP-11-015/trunk/TOP-11-015.tex $
\RCS$Id: TOP-11-015.tex 156083 2012-11-02 08:41:19Z stadie $
\newlength\cmsFigWidth
\ifthenelse{\boolean{cms@external}}{\setlength\cmsFigWidth{0.85\columnwidth}}{\setlength\cmsFigWidth{0.4\textwidth}}
\ifthenelse{\boolean{cms@external}}{\providecommand{\cmsLeft}{top}}{\providecommand{\cmsLeft}{left}}
\ifthenelse{\boolean{cms@external}}{\providecommand{\cmsRight}{bottom}}{\providecommand{\cmsRight}{right}}
\newcommand{\mtop}{\ensuremath{m_\cPqt}\xspace}
\cmsNoteHeader{TOP-11-015} 
\title{Measurement of the top-quark mass in $\ttbar$ events with lepton+jets final states in $\Pp\Pp$ collisions at $\sqrt{s}=7$\TeV}

\date{\today}

\abstract{
The mass of the top quark is measured using a sample of \ttbar candidate events with one electron or muon and at least four jets in the final state, collected by CMS in $\Pp\Pp$ collisions at $\sqrt{s}=7$\TeV at the LHC.
A total of 5174 candidate events is selected from data corresponding to an integrated luminosity of 5.0\fbinv. For each event the mass is reconstructed from a kinematic fit of the decay products to a \ttbar hypothesis. The top-quark mass is determined  simultaneously with the jet energy scale (JES), constrained by the known mass of the \PW\ boson in $\Pq\Paq$ decays, to be $173.49\pm0.43\,\text{(stat.+JES)}\pm0.98\,\text{(syst.)}\GeV$.
}

\hypersetup{%
pdfauthor={CMS Collaboration},%
pdftitle={Measurement of the top-quark mass in t t-bar events with lepton+jets final states in pp collisions at sqrt(s)=7 TeV},%
pdfsubject={CMS},%
pdfkeywords={CMS, physics}}

\maketitle 

\section{Introduction}
\label{sec:introduction}
The top-quark mass ($\mtop$) is an essential parameter of the standard model. Its measurement also provides an important benchmark for the performance and calibration of the Compact Muon Solenoid (CMS) detector~\cite{CMS:2008zzk}.
The top-quark mass has been determined  with a high precision at the Fermilab Tevatron to be $\mtop=173.18\pm0.94\GeV$ from \ttbar events in $\Pp\Pap$ collisions~\cite{Lancaster:2011wr}.
Measurements have been performed in different top-quark decay channels and using different methods, with the most precise single measurement of $\mtop=172.85\pm1.11$\GeV being from  the CDF Collaboration in the lepton+jets channel using a template method~\cite{Aaltonen:2012zzz}.
At the CERN Large Hadron Collider (LHC), the CMS Collaboration has measured the top-quark mass in the dilepton channel~\cite{Chatrchyan:2011nb,CMS-TOP-11-016-001}, where the latest measurement with an integrated luminosity of 5.0\fbinv yields  $\mtop=172.5\pm1.5$\GeV~\cite{CMS-TOP-11-016-001}. The ATLAS Collaboration has published a measurement in the  lepton+jets channel with an integrated luminosity of 1.04\fbinv, also using a template method, that  gives $\mtop=174.5\pm2.4$\GeV~\cite{Aad:2012aa}.

In the analysis presented here, we select events containing top-quark pairs where each top quark decays weakly via $\cPqt \rightarrow \cPqb\PW$, with one \PW\ boson decaying into a charged lepton and its neutrino, and the other into a quark-antiquark ($\cPq\cPaq$) pair.
Hence, the final state consists of a lepton, four jets, and an undetected neutrino.
The analysis employs a kinematic fit of the  decay products to a \ttbar hypothesis and two-dimensional (2D) likelihood functions for each event to estimate simultaneously both the top-quark mass and the jet energy scale (JES).
The invariant mass of the two jets associated with the  $\PW  \rightarrow \cPq\cPaq$ decay serves as an additional observable in the likelihood functions to estimate the JES directly exploiting the precise knowledge of the W-boson mass from previous measurements~\cite{pdg}.

\section{The CMS detector}
\label{sec:detector}
The central feature of the CMS apparatus is a superconducting solenoid, of 6\unit{m} internal diameter, providing a field of 3.8\unit{T}.
Within the field volume are a silicon pixel and strip tracker, a crystal electromagnetic calorimeter (ECAL) and a brass/scintillator hadron calorimeter (HCAL).
Muons are measured in gas-ionization detectors embedded in the steel return yoke.

CMS uses a right-handed coordinate system, with the origin at the nominal interaction point, the $x$ axis pointing to the center of the LHC ring, the $y$ axis pointing up (perpendicular to the plane of the LHC ring), and the $z$ axis along the counterclockwise-beam direction.
The polar angle, $\theta$, is measured from the positive $z$ axis and the azimuthal angle, $\phi$, is measured in the $x$-$y$ plane.

The CMS tracker consists of  silicon pixel and silicon strip detector modules,
covering the pseudorapidity range $|\eta|< 2.5$, where $\eta = -\ln[\tan(\theta/2)]$. The ECAL uses lead-tungstate crystals as scintillating material
and provides coverage for pseudorapidity $\vert \eta \vert< 1.5$ in
the central barrel region and $1.5 <\vert \eta \vert < 3.0$ in the forward endcap regions. Preshower detectors are installed in front of the endcaps to identify $\Pgpz$ mesons.
The HCAL consists of a set of sampling calorimeters, that utilize alternating layers of brass as absorber and plastic scintillator as active material.
In addition to the barrel and endcap detectors, CMS has extensive forward calorimetry that extends the coverage to $|\eta| < 5$.
The muon system includes barrel drift tubes covering the pseudorapidity range $|\eta|< 1.2$, endcap
cathode strip chambers ($0.9< |\eta|< 2.5$), and resistive plate chambers ($|\eta|< 1.6$).
A two-tier trigger system selects the most interesting pp collision
events for use  in physics analyses.
A more detailed description of the CMS detector can be found in Ref.~\cite{CMS:2008zzk}.

\section{Data sample and event selection}
\label{sec:selection}
The full 2011 data sample has been analyzed, corresponding to an integrated luminosity of $5.0\pm0.1$\fbinv at $\sqrt{s} = 7$\TeV.
Events are required to pass a single-muon trigger or an electron+jets trigger. The minimum trigger threshold on the  transverse momentum (\pt) of an isolated muon ranges from $17$\GeV to $24$\GeV, depending on the instantaneous pp luminosity.  The electron+jets trigger requires one isolated electron with $\pt>25$\GeV and at least three jets with $\pt>30$\GeV.

We use simulated events to develop and evaluate the analysis method. The \ttbar signal and W/Z+jets background events have been generated
with the \MADGRAPH~5.1.1.0 matrix element generator~\cite{Alwall:2011uj},
\PYTHIA~6.424 parton showering~\cite{Sjostrand:2006za} using the $Z2$ tune~\cite{Chatrchyan:2011id}, and a full simulation of the CMS detector based on
\GEANTfour~\cite{Agostinelli:2002hh}.
For the \ttbar signal events, nine different top-quark mass values ranging from 161.5\GeV to 184.5\GeV have been assumed.
The single-top-quark background has been simulated using \POWHEG 301~\cite{Alioli:2009je,Re:2010bp,Nason:2004rx,Frixione:2007vw,Alioli:2010xd}, assuming a top-quark mass of $172.5$\GeV.
The \ttbar, W/Z+jets, and single-top-quark samples are normalized to the theoretical predictions described in Refs.~\cite{Kidonakis:2010dk,Melnikov:2006kv,Campbell:2010ff}.
The simulation includes effects of additional overlapping minimum-bias events (pileup) that match their distribution in data.
Furthermore, the jet energy resolution in simulation is scaled to match the resolution observed in data~\cite{Chatrchyan:2011ds}.

Events are reconstructed with the particle-flow (PF) algorithm~\cite{CMS-PAS-PFT-10-002} that combines the information from all CMS sub-detectors to identify and reconstruct individual objects produced in the pp collision.
The reconstructed particles include muons, electrons, photons, charged hadrons, and neutral hadrons.
Charged particles are required to originate from the primary collision vertex, identified as the reconstructed vertex with the largest value of $\Sigma \pt^2$ for its associated tracks.
The list of charged and neutral PF particles is used as input for jet clustering based on the anti-\kt algorithm with a distance parameter of 0.5~\cite{Cacciari:2005hq,Cacciari:2008gp}.
Particles identified as isolated muons and electrons are excluded from the clustering.
The momentum of a jet is determined from the vector sum of all particle momenta in the jet. From simulation, the reconstructed jet momentum is found to be typically within 5--10\% of the true jet momentum.
Jet energy corrections are applied to all the jets in data and simulation~\cite{Chatrchyan:2011ds}.
These corrections are defined as a function of the transverse momentum density of an event~\cite{Cacciari:2007fd,Cacciari:2008gn,Cacciari:2011ma}, and also depend on the \pt and $\eta$ of the reconstructed jet. This procedure provides a uniform energy response at the particle level with a low dependency on pileup.
An additional residual correction, measured from the momentum imbalance observed in dijet and photon+jet/Z+jet events, is also applied to the jets in data.
Finally, the missing transverse momentum is given by the negative vector sum of the transverse momenta of all particles found by the PF algorithm.

Events are selected to have exactly one isolated lepton ($\ell$), muon or electron, with $\pt>30$\GeV and $| \eta |<2.1$ and at least four jets with $\pt>30$\GeV and  $| \eta |<2.4$.
In addition, jets originating from bottom (b) quarks are tagged with an algorithm that combines reconstructed secondary vertices and track-based lifetime information, the Combined Secondary Vertex Medium (CSVM) b tagger described in Ref.~\cite{CMS-PAS-BTV-11-004}.
We require at least two b-tagged jets and select 17\,985 \ttbar candidate events in data. The estimated selection efficiency for \ttbar signal is $2.3\%$.
From simulation, the event composition is expected to be  $90$\% \ttbar, $4$\% single top quark, $3$\% W+jets, and $3$\% other processes which include the multijet background. Hence, the selection leads to a very clean sample of \ttbar events (see also Refs.~\cite{Chatrchyan:2011yy,Chatrchyan:2012ub}).
\section{Kinematic fit}
\label{sec:kinematicfit}
A kinematic fit is employed to check the compatibility
of an event with the \ttbar hypothesis and thereby improve the resolution
of the measured quantities.
The fit constrains the event to the hypothesis for the production of two heavy particles of equal mass, each one decaying to a W boson and a b quark.  As indicated above, one of the W bosons decays into a lepton-neutrino pair, while the other W boson decays into a quark-antiquark pair.
The reconstructed masses of the two W bosons are constrained in the fit to $80.4$\GeV~\cite{pdg}.
A comprehensive description of the algorithm and constraints on the fit is available in Ref.~\cite{Abbott:1998dc}.

The inputs to the fitter are the four-momenta of the lepton and the four
leading jets, the missing transverse momentum, and their respective
resolutions.
The two b-tagged jets are candidates for the b quarks in the \ttbar hypothesis, while the two untagged jets serve as candidates for the light quarks for one of the W-boson decays. This leads to two possible parton-jet assignments per event.
For simulated \ttbar events, the parton-jet assignments can be classified as \emph{correct permutations} ($\mathrm{cp}$), \emph{wrong permutations} ($\mathrm{wp}$), and \emph{unmatched permutations} ($\mathrm{un}$), where, in the latter, at least one quark from the \ttbar decay is not matched to any of the four selected jets.

To increase the fraction of correct permutations, we require the goodness-of-fit (gof) probability for the kinematic fit with two degrees of freedom $P_\mathrm{gof} = P\left(\chi^2\right) = \exp\left(-\frac{1}{2} \chi^{2}\right)$ to be at least 0.2. This selects 2906 muon+jets and 2268 electron+jets events for the mass measurement.
For each event we use all permutations that fulfill this requirement, and weight the permutations by their $P_\mathrm{gof}$ values.
In simulation, the fraction of  correct permutations improves from $13$\% to $44$\% and the non-\ttbar background is reduced to $4\%$.

\begin{figure}
\noindent \begin{centering}
\subfloat[]{\noindent \begin{centering}
\includegraphics[width=0.5\textwidth]{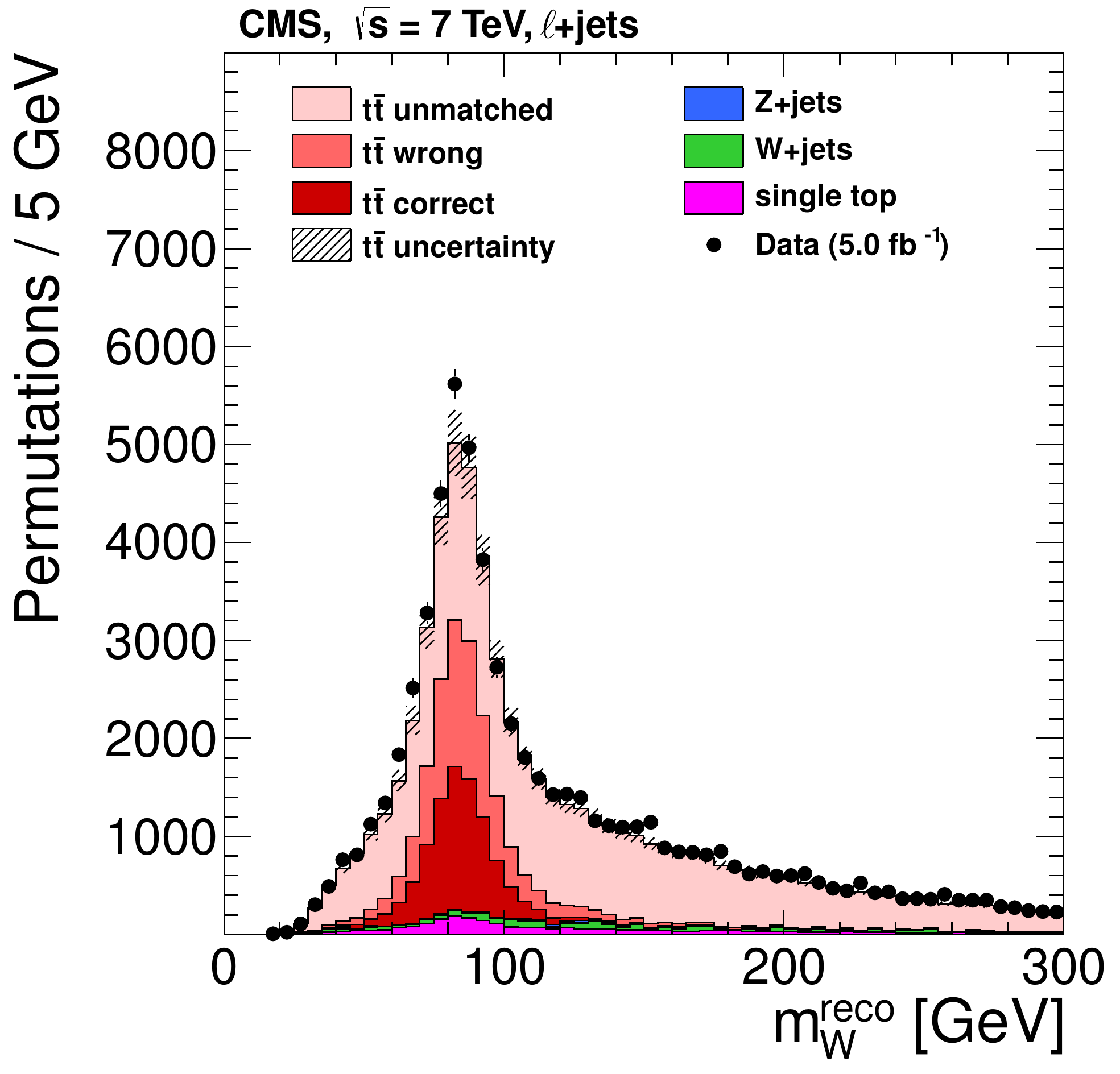}
\par\end{centering}
}
\subfloat[]{\noindent \begin{centering}
\includegraphics[width=0.5\textwidth]{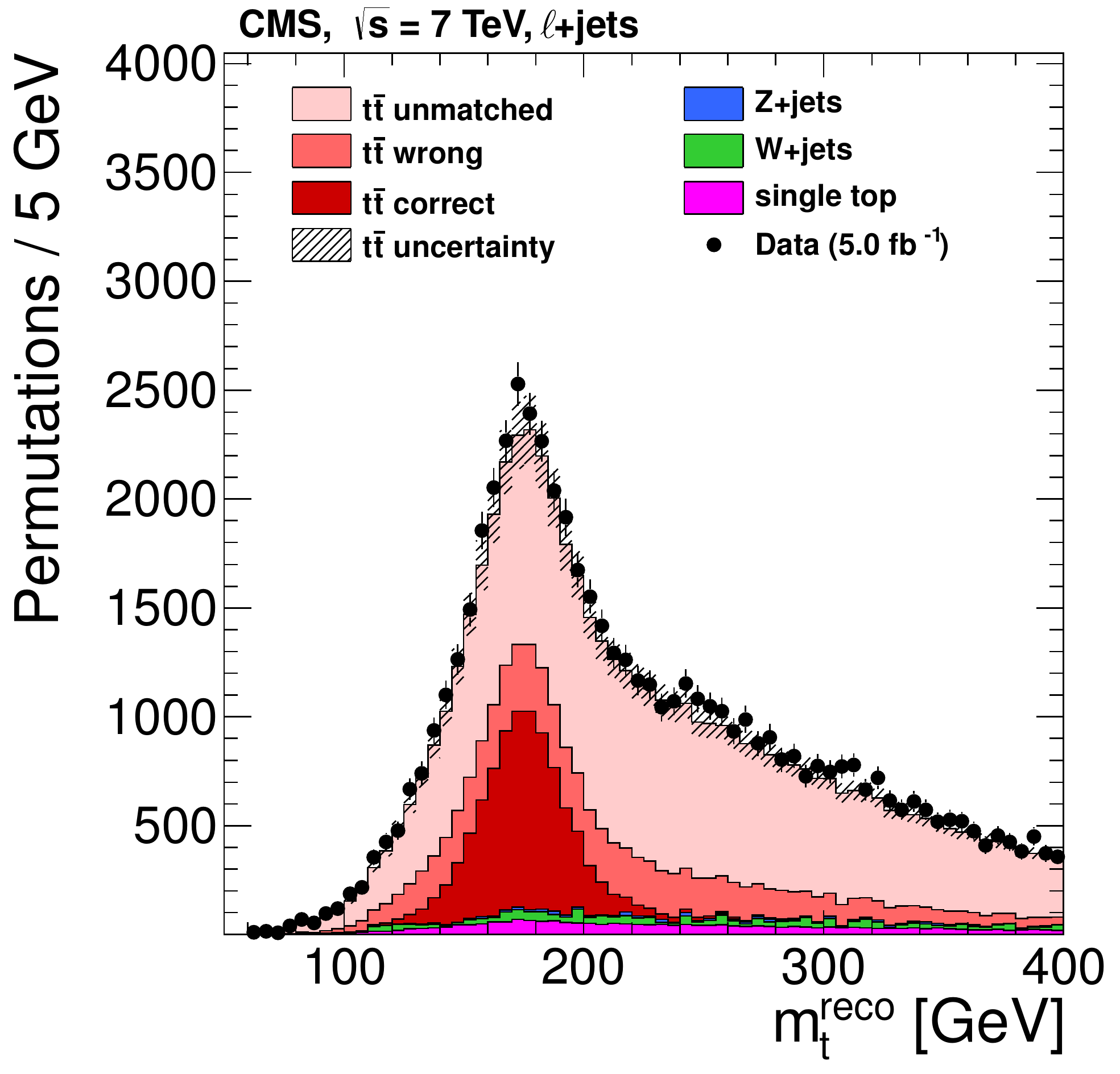}
\par\end{centering}
}
\\
\subfloat[]{\noindent \begin{centering}
\includegraphics[width=0.5\textwidth]{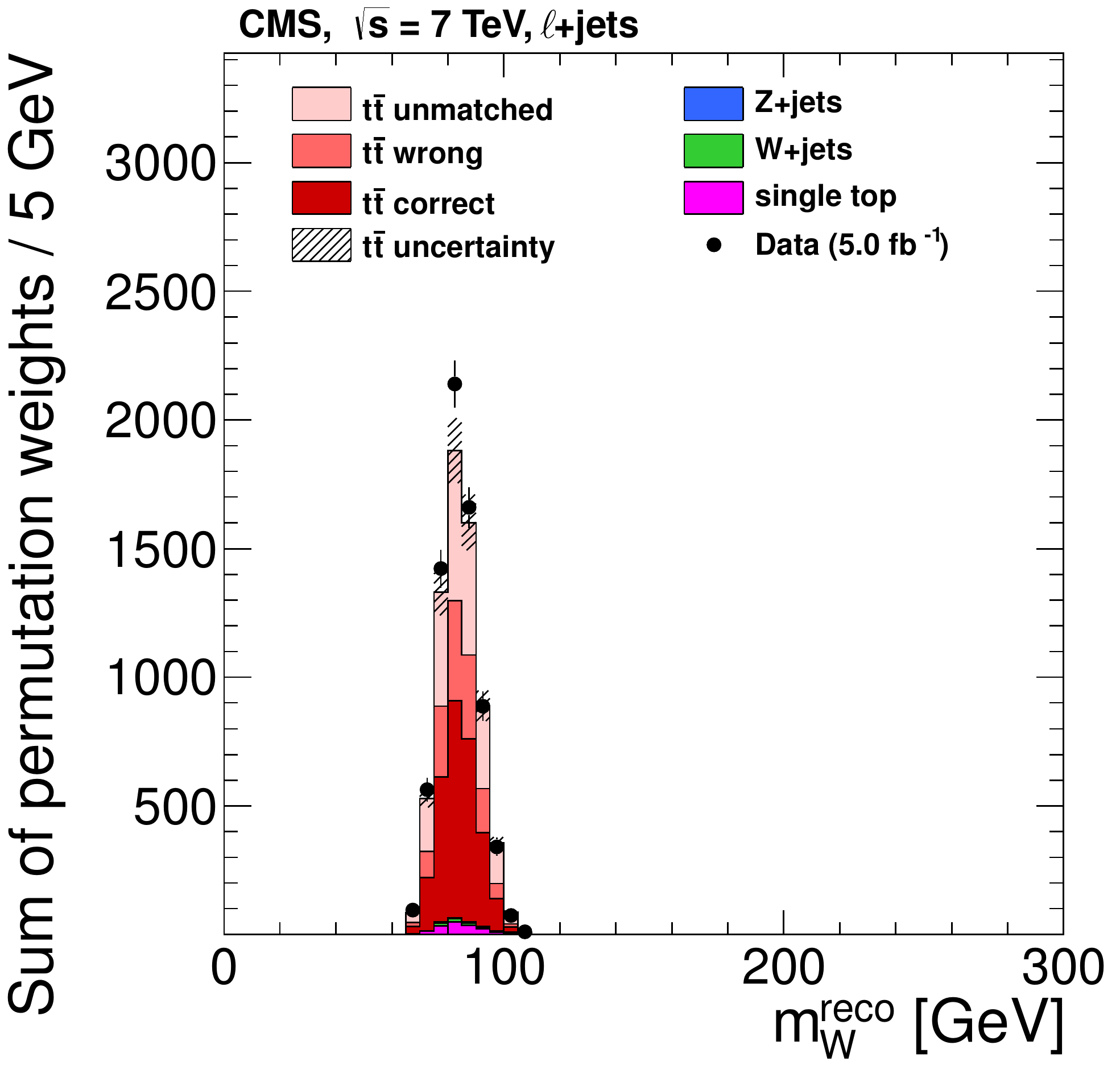}
\par\end{centering}
}
\subfloat[]{\noindent \begin{centering}
\includegraphics[width=0.5\textwidth]{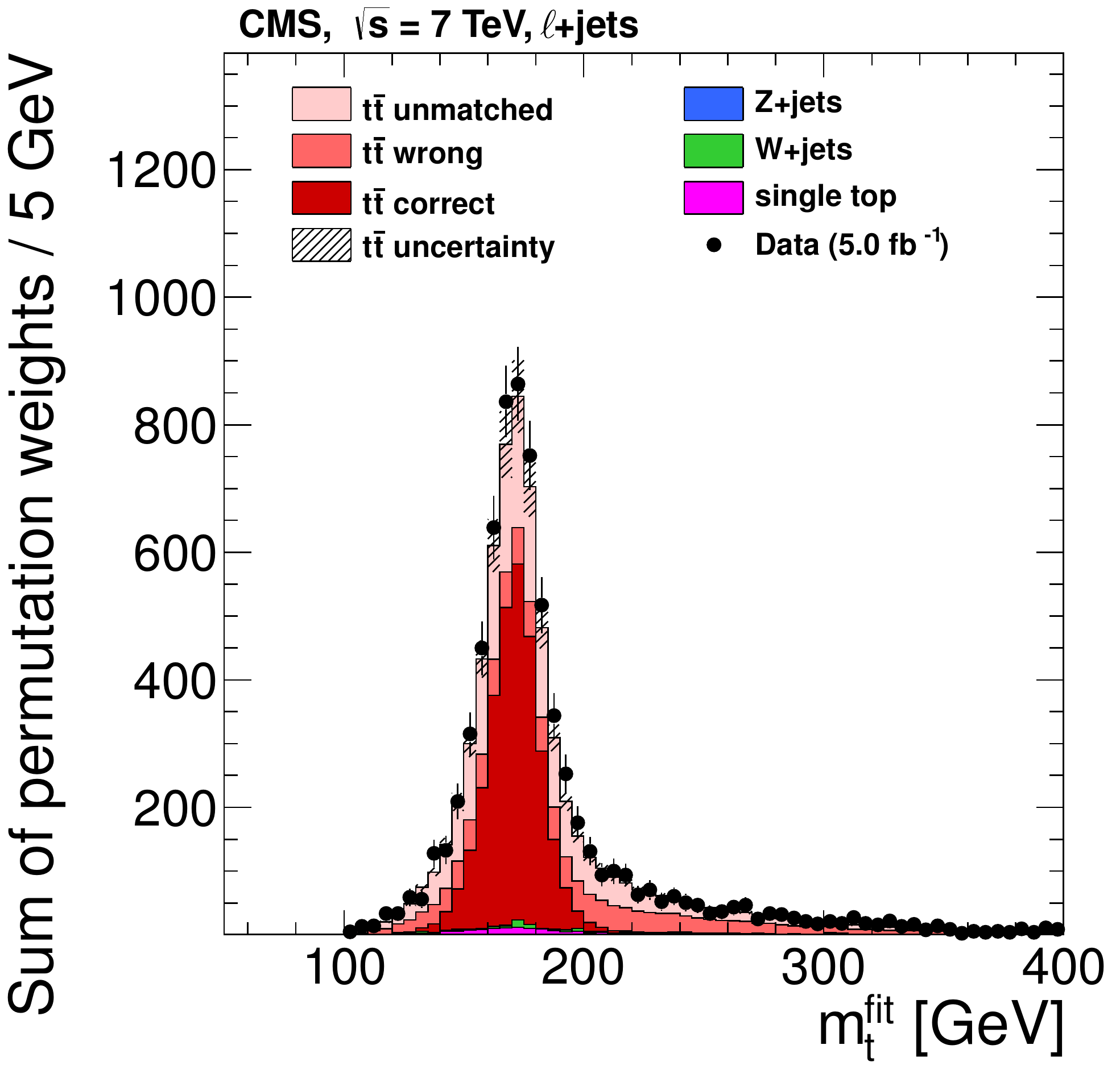}
\par\end{centering}
}
\par\end{centering}

\caption{\label{fig:controlplot-weighted-obs}
Reconstructed masses of (a) the W bosons decaying to $\cPq\cPaq$ pairs and (b) the corresponding top quarks, prior to the kinematic fitting to the \ttbar hypothesis. (c) and (d) show, respectively, the reconstructed W-boson masses and the fitted top-quark masses after the goodness-of-fit selection and the weighting by $P_\mathrm{gof}$. The distributions are normalized to the theoretical predictions described in Refs.~\cite{Kidonakis:2010dk,Melnikov:2006kv,Campbell:2010ff}.
The uncertainty on the predicted \ttbar cross section is indicated by the hatched area. The top-quark mass assumed in the simulation is $172.5$\GeV.}
\end{figure}

Figures~\ref{fig:controlplot-weighted-obs} (a) and  \ref{fig:controlplot-weighted-obs} (b) show, respectively, the distributions in the reconstructed mass $m_\PW^\text{reco}$ of the W boson decaying to a $\cPq\cPaq$ pair and the mass $\mtop^\text{reco}$ of the corresponding top quark for all possible permutations before the kinematic fit.
The reconstructed W-boson mass  $m_\PW^\text{reco}$ and the top-quark mass from the kinematic fit $\mtop^\text{fit}$,  after the $P_{\rm gof}$ selection and weighting, are displayed in Figs.~\ref{fig:controlplot-weighted-obs} (c)  and \ref{fig:controlplot-weighted-obs} (d).

\section{Ideogram method}
\label{sec:ideogram}
As the jet energy scale was found to be the leading systematic uncertainty in previous measurements of $\mtop$ in this channel, we chose to determine the JES and the top-quark mass simultaneously in a joint likelihood fit to the selected events.
The observable used for measuring $\mtop$ is the mass $\mtop^\text{fit}$ found in the kinematic fit. We take the reconstructed W-boson
mass $m_\PW^\text{reco}$, before it is constrained by the kinematic fit,
as an estimator for measuring \emph{in situ} a residual JES to be applied in addition to the CMS jet energy corrections described in Section~\ref{sec:selection}. This is in contrast to a similar measurement from the \DZERO Collaboration~\cite{Abazov:2007rk}, where the kinematic fit is performed for different JES hypotheses.

The ideogram method has been used by the DELPHI Collaboration to measure
the W-boson mass at the CERN LEP collider~\cite{Abdallah:2008xh}, and, at the Fermilab Tevatron collider, by the \DZERO Collaboration to measure the top-quark mass in the lepton+jets channel~\cite{Abazov:2007rk} and by the CDF Collaboration to measure the top-quark mass in the all-jet channel~\cite{Aaltonen:2006xc}.
The likelihood in the ideogram method is evaluated from
analytic expressions obtained and calibrated using simulated events.
This makes the ideogram method relatively fast and reliable.

The likelihood used to estimate the top-quark mass and the JES, given the observed data, can be defined as:
\begin{eqnarray*}
\mathcal{L}\left(\mtop,\mathrm{JES}|\text{sample}\right) & \propto & \mathcal{L}\left(\text{sample}|\mtop,\mathrm{JES}\right) =  \prod_\text{events}  \mathcal{L}\left(\text{event}|\mtop,\mathrm{JES}\right)^{w_\text{event}}\\
 & = & \prod_\text{events}\left(\sum_{i=1}^{n} c\,P_\mathrm{gof}\left(i\right)P\left(m_{\cPqt,i}^\text{fit},m_{\PW,i}^\text{reco}|\mtop,\mathrm{JES}\right)\right)^{w_\text{event}},
\end{eqnarray*}
where $\mtop$ and JES are the parameters to be determined, $n$ denotes the number of permutations in each event, and $c$ is a normalization constant.
We note that the contribution from background is not included in this expression, as the impact of background is found to be negligible after implementing the final selections described in Section~\ref{sec:kinematicfit}.
The ad hoc event weight $w_\text{event}=\sum_{i=1}^{n}c\,P_\mathrm{gof}\left(i\right)$
is introduced to reduce the impact of events without correct permutations. The sum of all event weights is normalized by $c$ to the total number of events.

\begin{figure}
\subfloat[]{\includegraphics[width=0.33\textwidth]{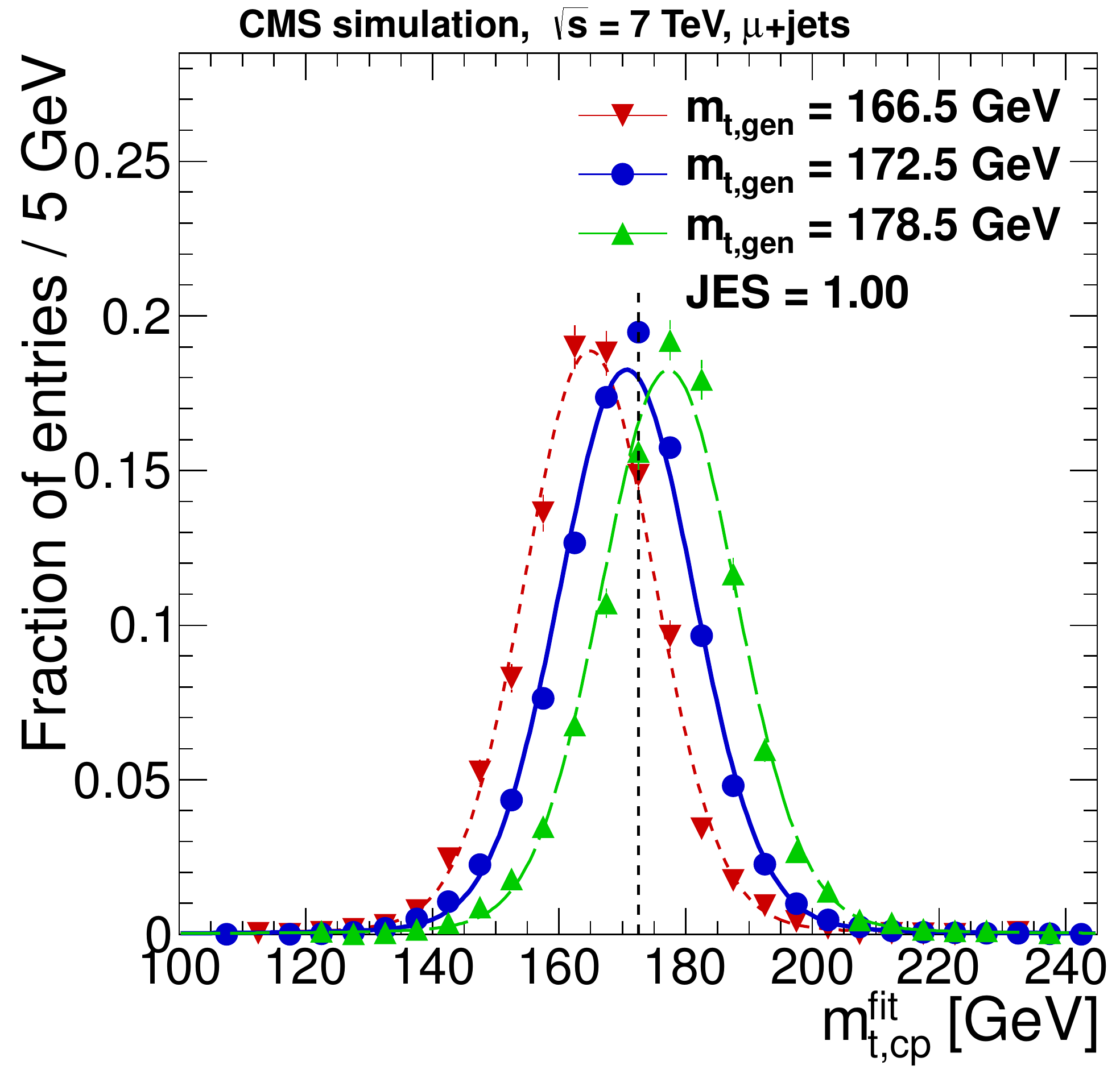}

}\subfloat[]{\includegraphics[width=0.33\textwidth]{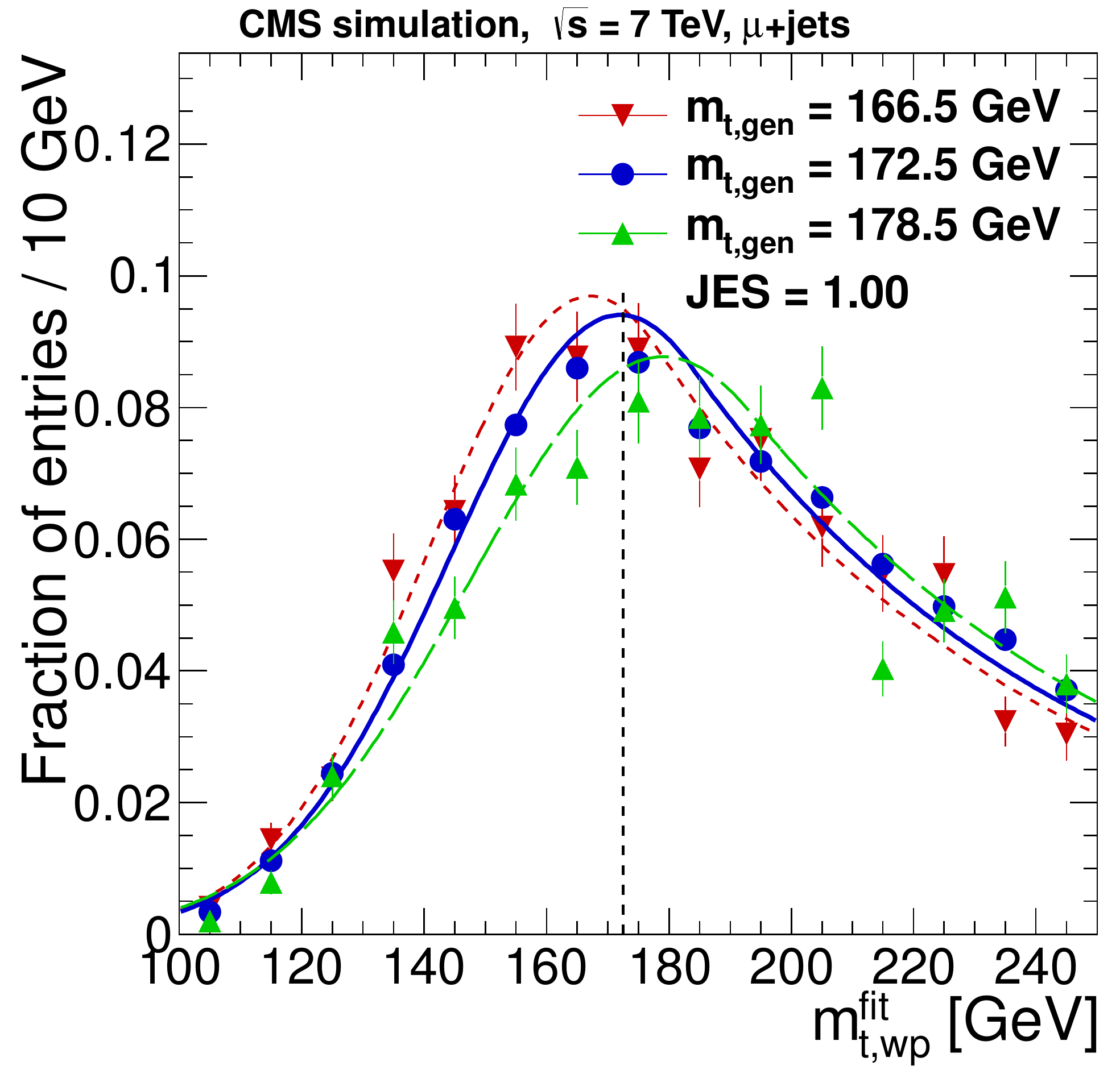}

}\subfloat[]{\includegraphics[width=0.33\textwidth]{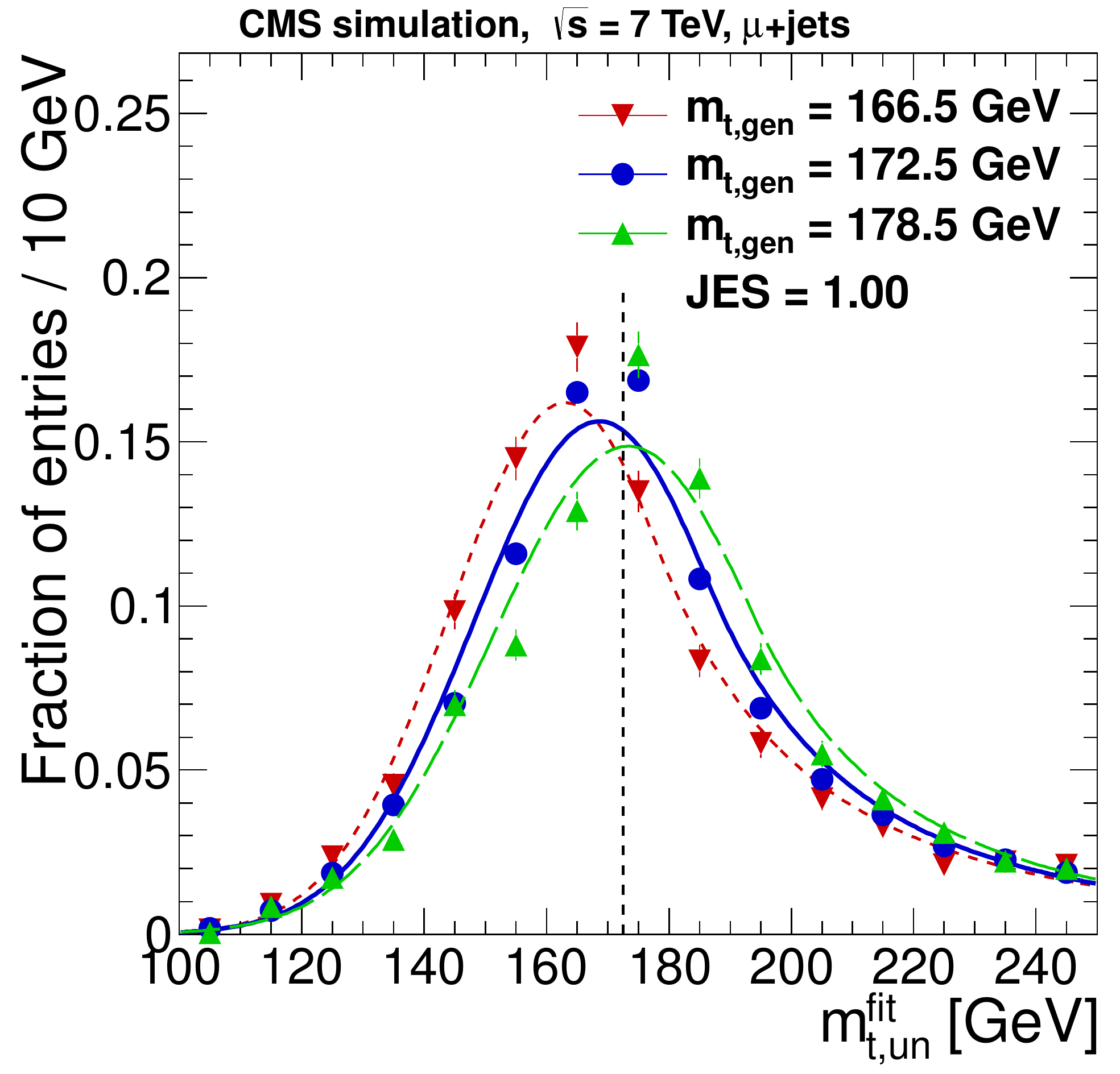}

}

\subfloat[]{\includegraphics[width=0.33\textwidth]{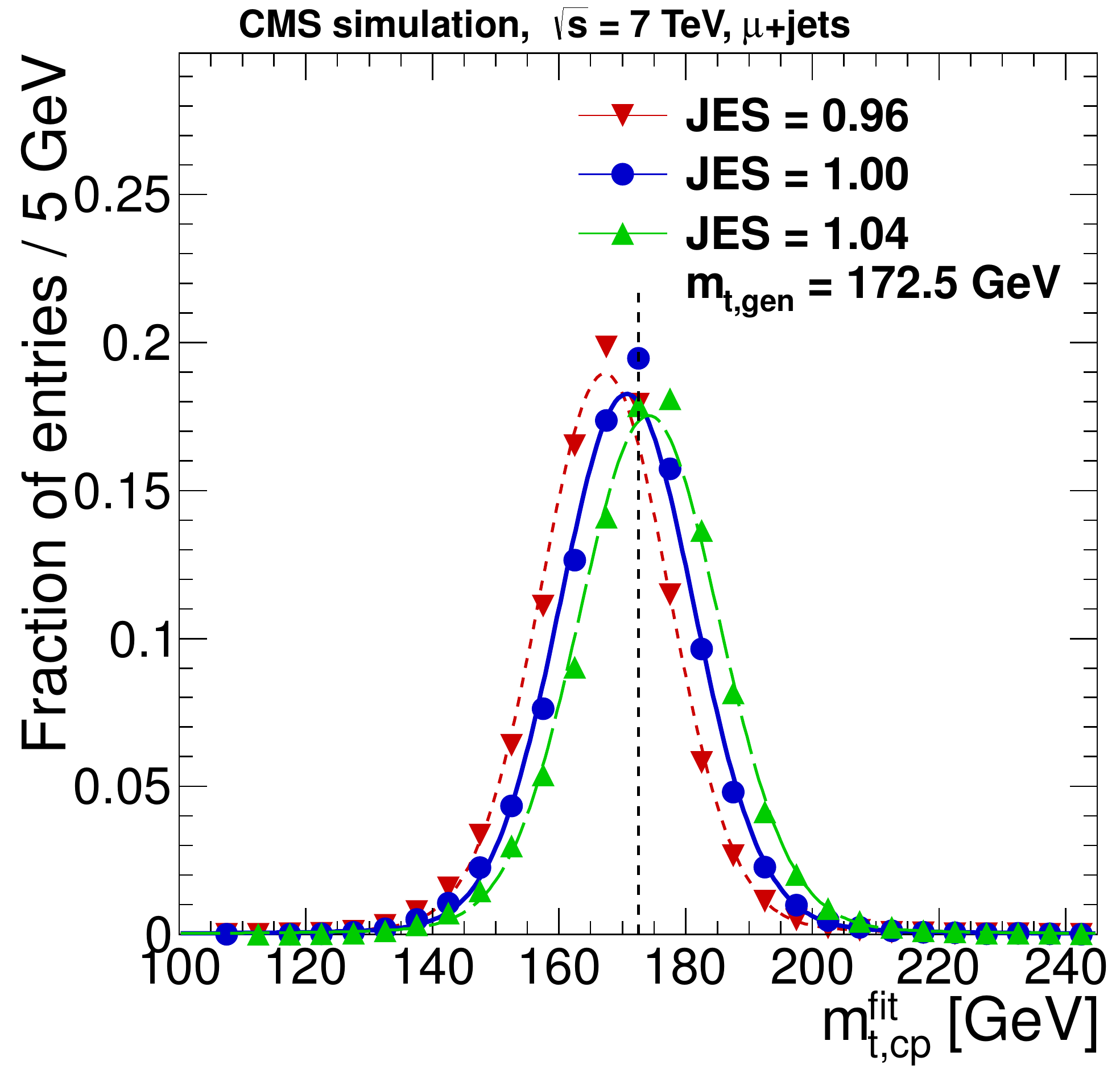}

}\subfloat[]{\includegraphics[width=0.33\textwidth]{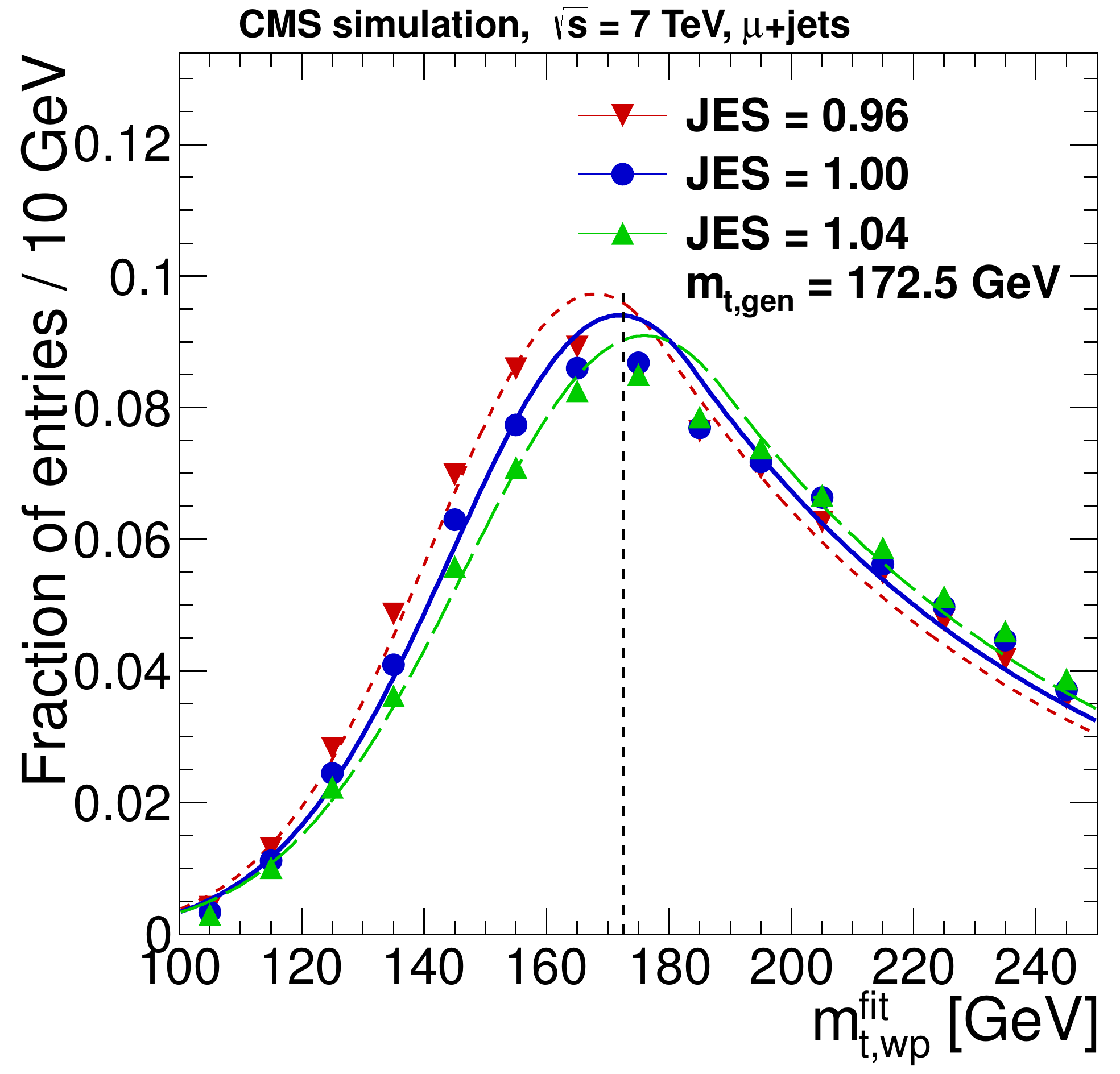}

}\subfloat[]{\includegraphics[width=0.33\textwidth]{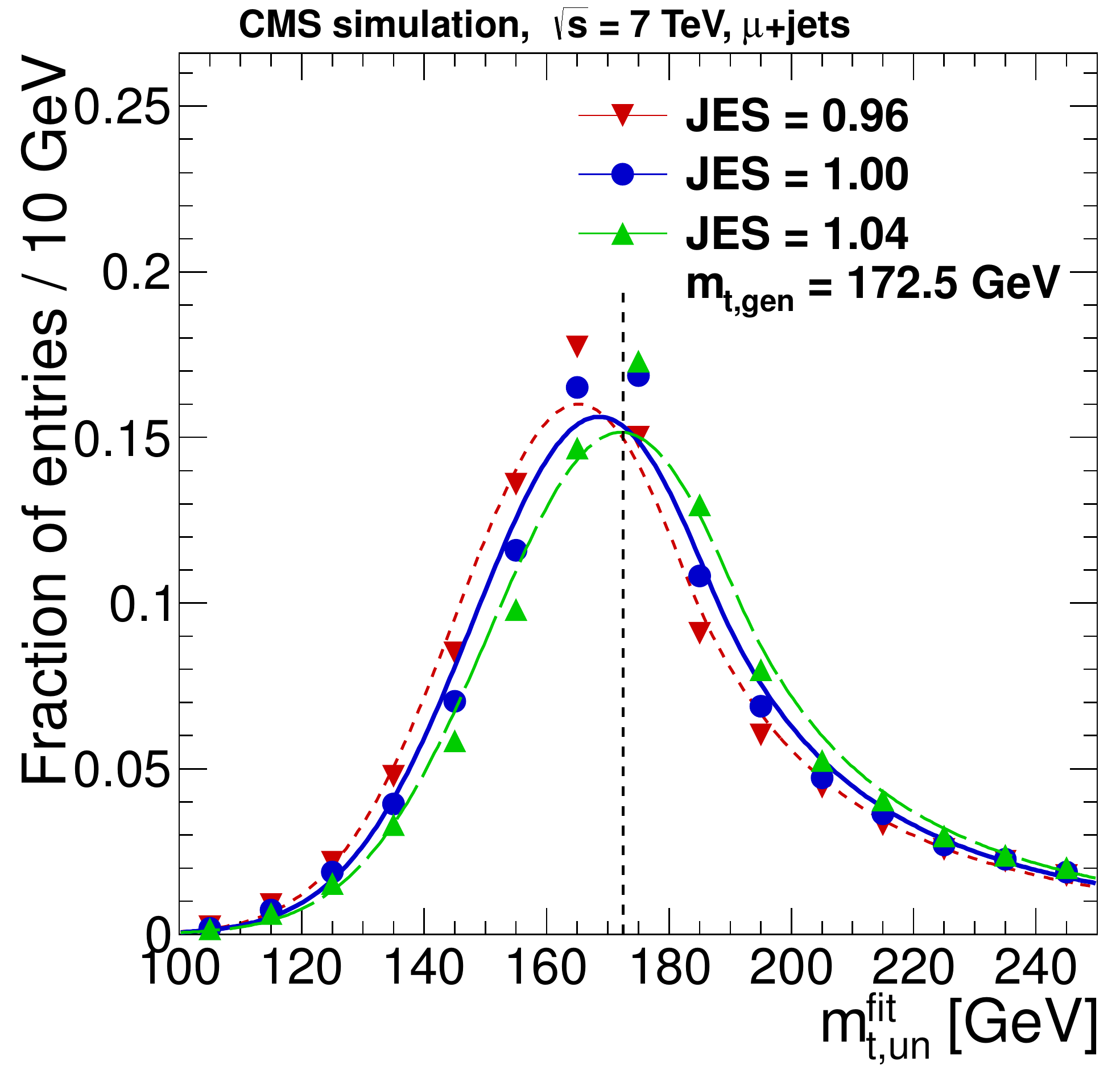}

}

\subfloat[]{\includegraphics[width=0.33\textwidth]{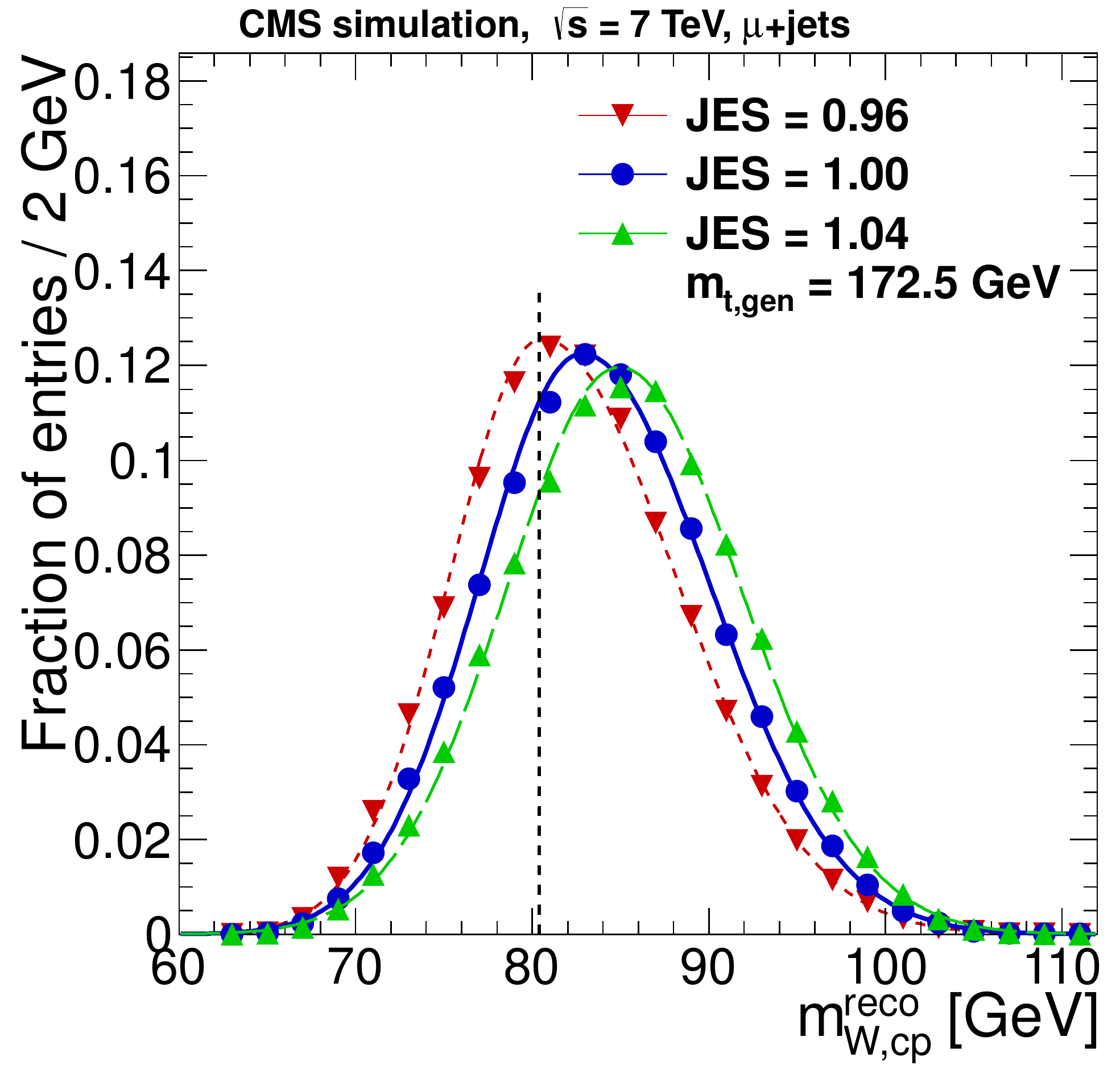}

}\subfloat[]{\includegraphics[width=0.33\textwidth]{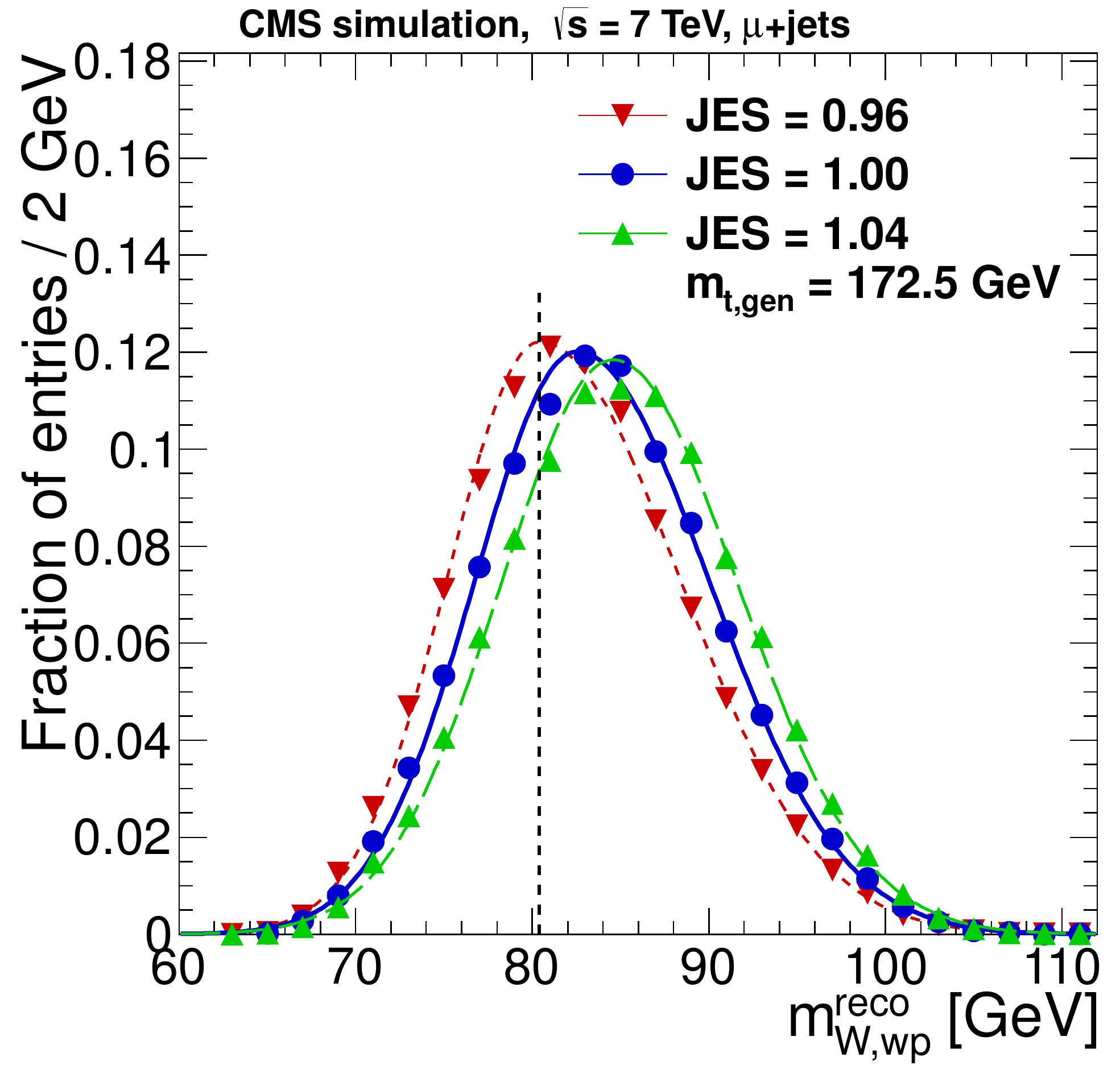}

}\subfloat[]{\includegraphics[width=0.33\textwidth]{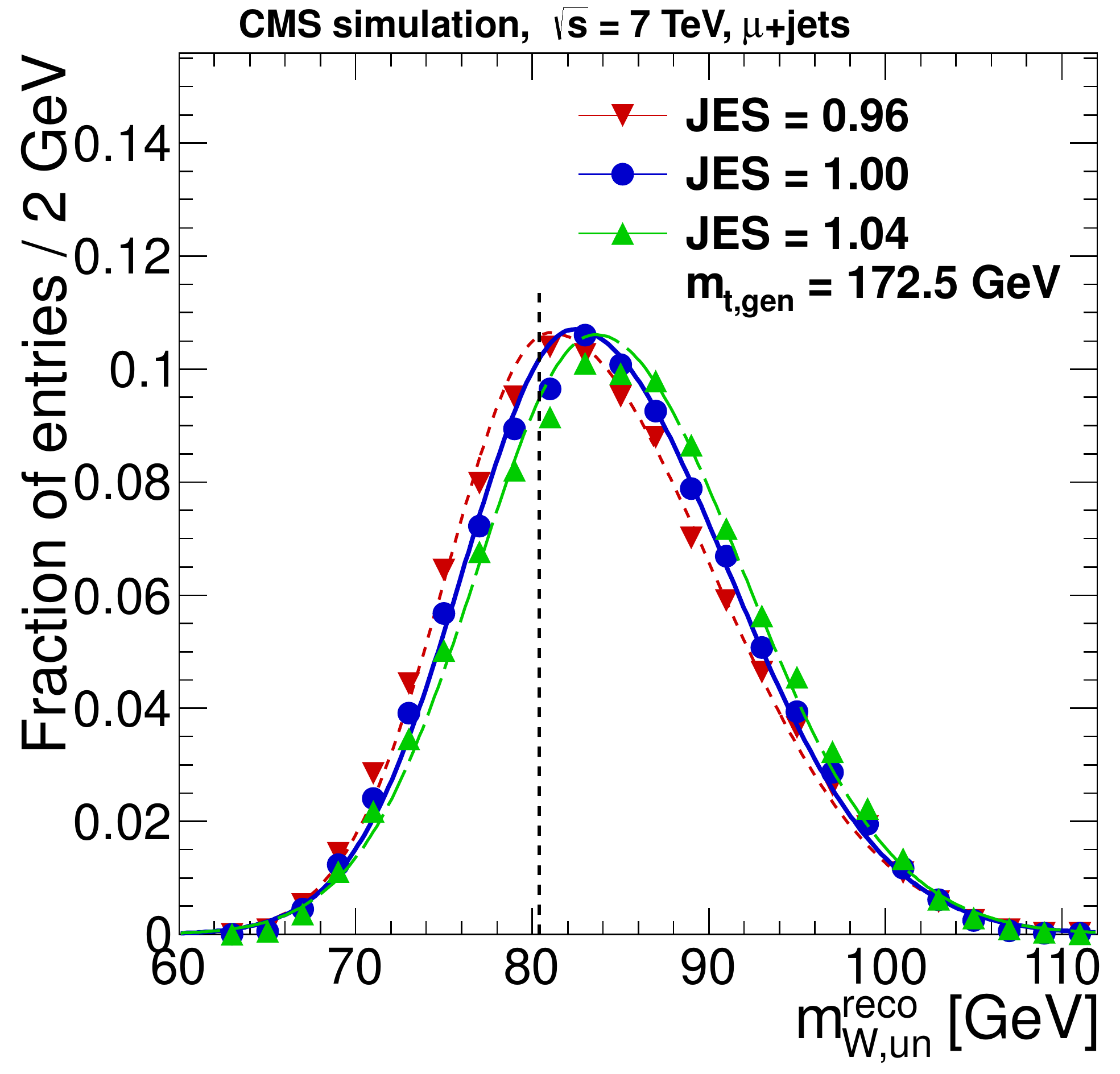}

}\caption{\label{fig:shapes} Simulated $\mtop^\text{fit}$ distributions of (a,d) correct,
(b,e) wrong, and (c,f) unmatched \ttbar permutations, for three
generated masses $m_{\cPqt,\text{gen}}$ with $\mathrm{JES} =1$ in (a), (b) and (c), and for three jet energy scales with $m_{\cPqt,\text{gen}} = 172.5$\GeV in (d), (e) and (f). The vertical dashed line corresponds to $\mtop^\text{fit} = 172.5$\GeV.
The $m_\PW ^\text{reco}$ distributions are shown for (g) correct,
(h) wrong, and (i) unmatched \ttbar permutations for three jet energy scales with $m_{\cPqt,\text{gen}} = 172.5$\GeV. The vertical dashed lines in (g), (h) and (i) indicate the accepted value of the W-boson mass of $80.4$\GeV. All distributions are shown for the muon+jets channel.}
\end{figure}

Due to the mass constraint on the W boson in the kinematic fit, the correlation coefficient between  $\mtop^\text{fit}$  and $m_\PW^\text{reco}$ is only $0.026$ for simulated \ttbar events. Hence, $\mtop^\text{fit}$  and $m_\PW^\text{reco}$ can be treated as uncorrelated and the probability $P\left(m_{\cPqt,i}^\text{fit},m_{\PW,i}^\text{reco}|\mtop,\mathrm{JES}\right)$ for the permutation $i$ factorizes into:
\begin{eqnarray*}
P\Big(m_{\cPqt,i}^\text{fit},m_{\PW ,i}^\text{reco}|\mtop,\mathrm{JES}\Big) & = & \sum_{j}f_{j}P_{j}\Big(m_{\cPqt,i}^\text{fit}|\mtop,\mathrm{JES}\Big)\times P_{j}\Big(m_{\PW ,i}^\text{reco}|\mtop,\mathrm{JES}\Big),
\end{eqnarray*}
where $f_j$, with $j$ representing \emph{cp}, \emph{wp} or \emph{un}, is the relative fraction of the three kinds of permutations.
The relative fractions $f_j$ and probability density distributions $P_{j}$ are determined separately for the muon and electron channels from simulated \ttbar events generated for the nine different top-quark mass ($m_{\cPqt,\text{gen}}$) values and three different JES values (0.96, 1.00, and 1.04).
Each of the $\mtop^\text{fit}$ distributions is fitted either with a Breit-Wigner function, convoluted with a Gaussian resolution for correct permutations, or with a Crystal Ball function, for wrong and unmatched permutations, for different $m_{\cPqt,\text{gen}}$ and JES values.
The corresponding $m_\PW^\text{reco}$ distributions
are distorted by the $P_\mathrm{gof}$ requirement and weighting because permutations with reconstructed W-boson masses close to $80.4$\GeV are preferred in the kinematic fit.
The $m_\PW^\text{reco}$ distributions are therefore fitted with asymmetric
Gaussian functions.
Figure~\ref{fig:shapes} compares the $\mtop^\text{fit}$ and $m_\PW^\text{reco}$ distributions for the three different kinds of permutations and three choices of input top-quark mass and JES to the probability density distributions used for the muon channel. A similar behavior for $\mtop^\text{fit}$ and $m_\PW^\text{reco}$ is seen for the electron channel.
The parameters of all fitted functions are parameterized linearly
in terms of the generated top-quark mass, JES, and the product of the two.

We obtain separate likelihoods for the muon and electron channels and examine the product of these likelihoods to combine the channels.
The most likely top-quark mass and JES are obtained by minimizing
$-2\ln  \mathcal{L}\left(\mtop,\mathrm{JES}|\text{sample}\right)$.

\section{Calibration of the ideogram method}
\label{sec:calibration}
As the analysis method contains some simplifications, it has to be checked for
possible biases and for the correct estimation of the statistical uncertainty.
For each combination of the nine $m_{\cPqt,\text{gen}}$ values and the three $\mathrm{JES}$ scales, we conduct 10\,000 pseudo-experiments, separately for the muon and electron channels, using simulated \ttbar events. We extract $m_{\cPqt,\text{extr}}$ and $\mathrm{JES}_\text{extr}$ from each pseudo-experiment which corresponds to the same integrated luminosity as the one analyzed in data. This results in 27 calibration points in the ($\mtop,\mathrm{JES}$) plane.

\begin{figure}
\subfloat[]{\includegraphics[width=0.5\textwidth]{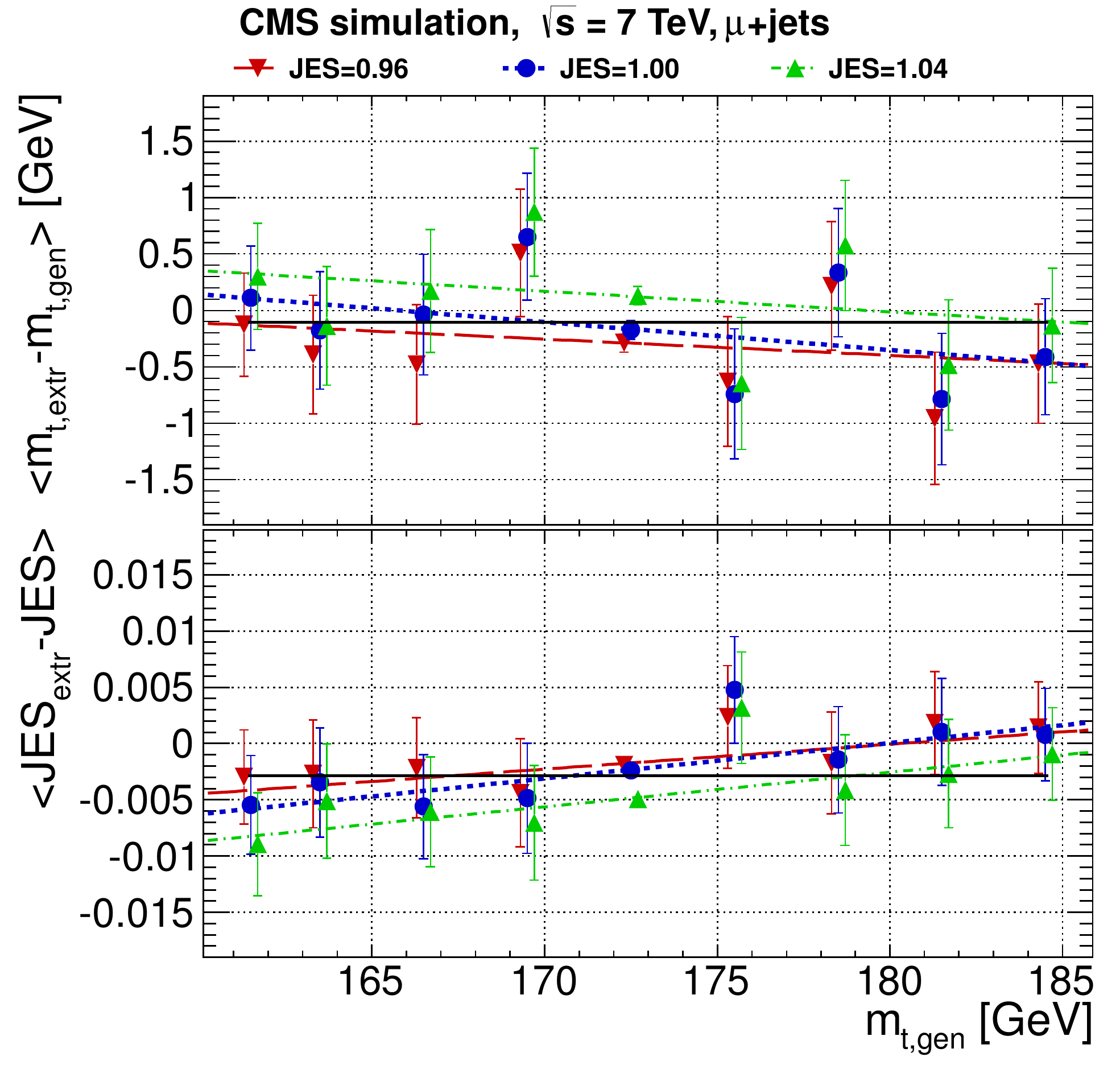}
}
\subfloat[]{\includegraphics[width=0.5\textwidth]{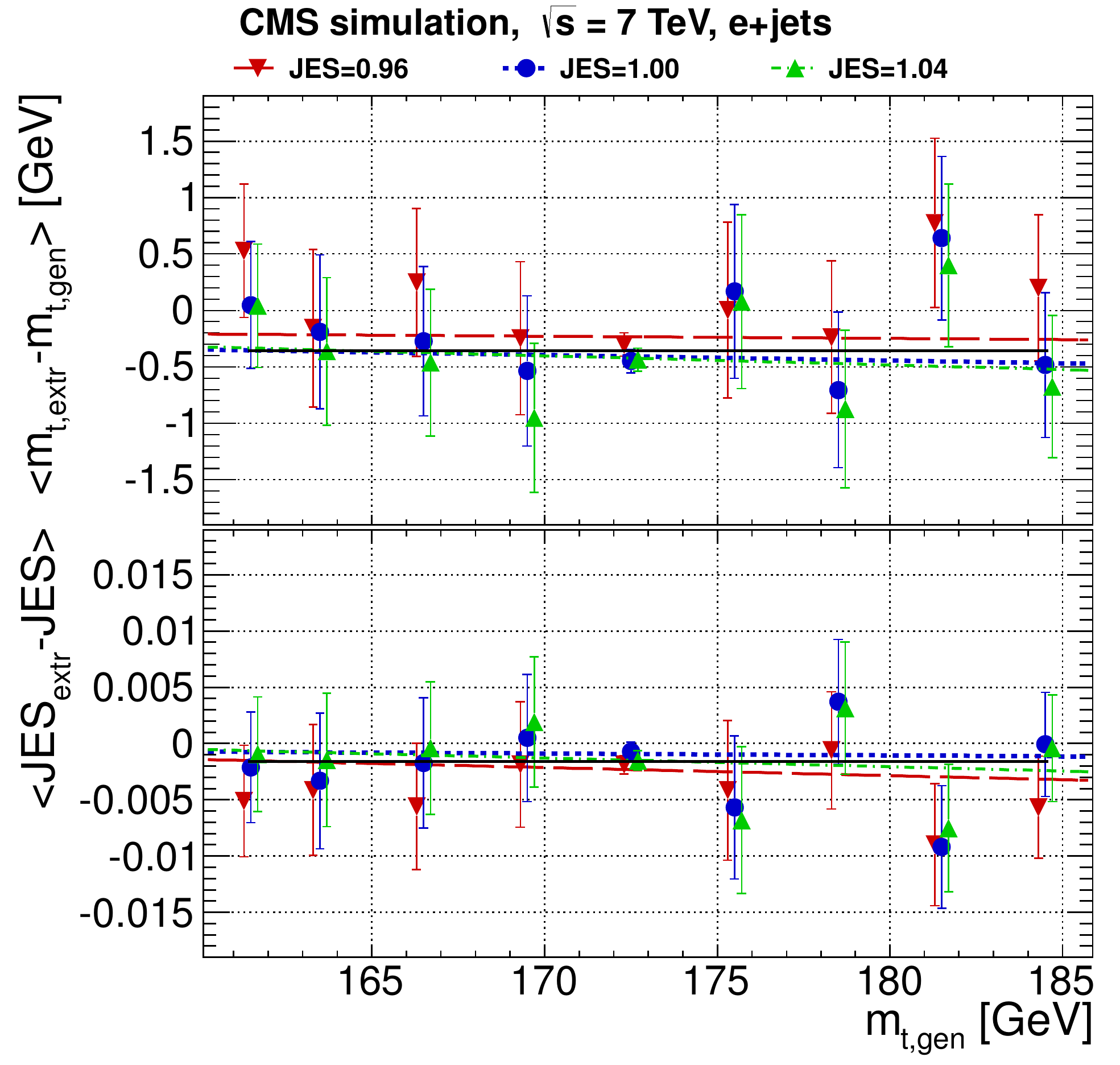}
}

\caption{\label{fig:Calibration} Mean difference between the extracted $ m_{\cPqt,\text{extr}}$ and each generated $m_{\cPqt,\text{gen}}$ and between $\mathrm{JES}_\text{extr}$ and $\mathrm{JES}_\text{gen}$ for the (a) muon channel and (b) electron channel, before the calibration, as a function of different generated $m_{\cPqt,\text{gen}}$ and three values of JES. The colored dashed lines correspond to straight line fits, which are used to correct the final likelihoods. The black solid line corresponds to an assumption of a constant calibration for all mass and JES points in each channel.
}
\end{figure}

The biases are defined as:
\begin{eqnarray*}
\mbox{mass bias} & = & \Big<m_{\cPqt,\text{extr}}-m_{\cPqt,\text{gen}}\Big>,\\
\mbox{JES bias} & = & \Big<\mathrm{JES}_\text{extr}-\mathrm{JES}\Big>.
\end{eqnarray*}
The biases in mass and in JES are shown in Fig.~\ref{fig:Calibration} as a function of $m_{\cPqt,\text{gen}}$ for the three values of ${\rm JES}$,
and are fitted to a linear dependence to each value of $\mathrm{JES}$.
A fit of the calibration points to a constant serves as a
quality estimator of the overall calibration.
Corrections for calibrating the top-quark mass $m_{\cPqt,\text{cal}}$ and
the jet energy scale  $\mathrm{JES}_\text{cal}$ are obtained from the fitted linear functions. These final corrections therefore depend linearly on the extracted values of top-quark mass, JES, and the product of the two.

\begin{figure}
\subfloat[]{\includegraphics[width=0.5\textwidth]{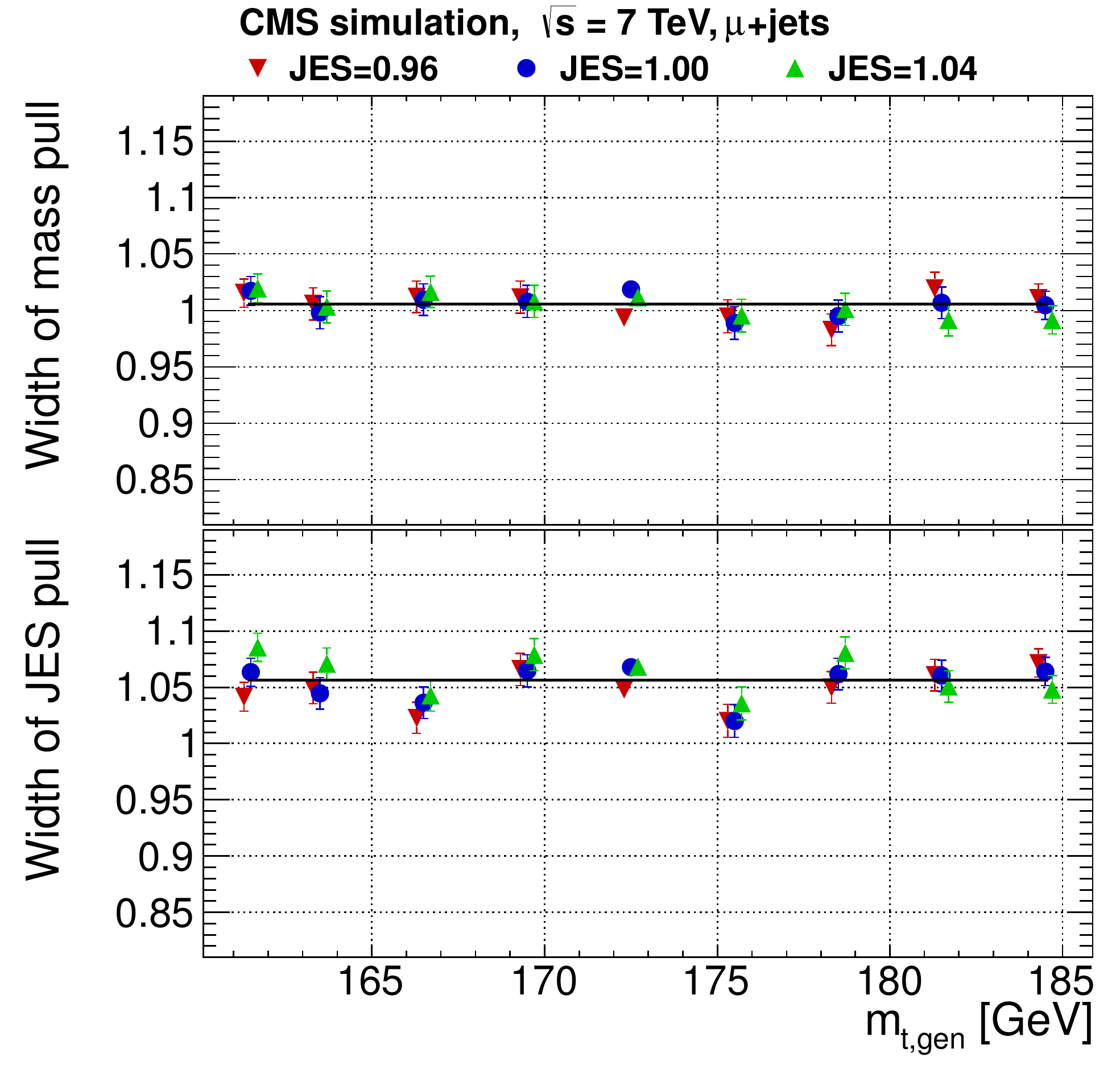}
}
\subfloat[]{\includegraphics[width=0.5\textwidth]{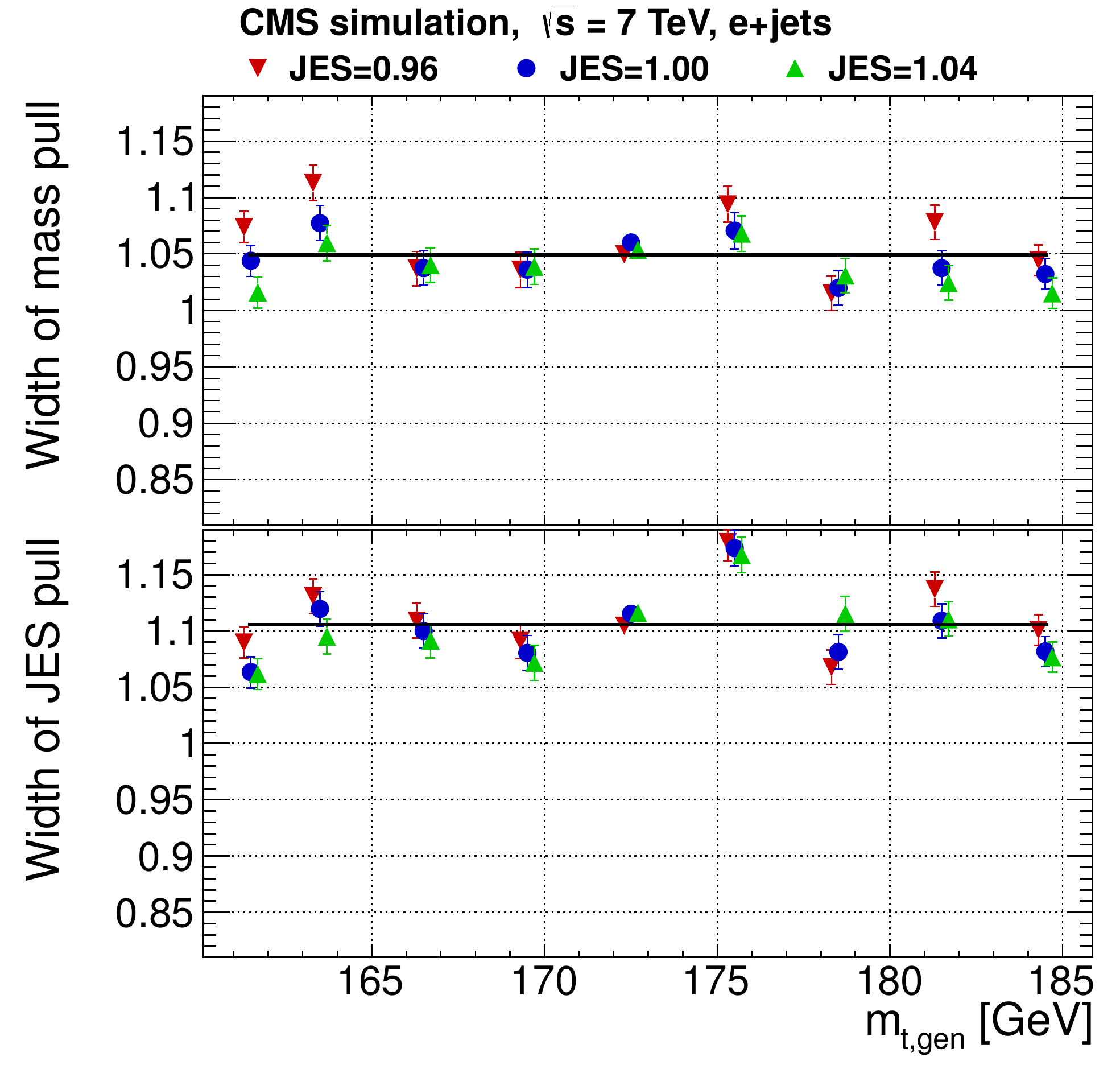}
}

\caption{\label{fig:Pull}  Width of the pull distribution for the calibrated measurement of $\mtop$ and JES as a function of different generated $m_{\rm t,gen}$ and three values of JES. The muon channel is shown in (a) and the electron channel in (b). The black solid lines correspond to fits of constants to all calibration points, assuming no dependence on $\mtop$ or JES.
}
\end{figure}

Using pseudo-experiments with the calibrated likelihood, we fit a Gaussian function to the distribution of the pulls defined as:
\[
\mbox{pull}=\frac{m_{\cPqt,\text{cal}}-m_{\cPqt,\text{gen}}}{\sigma\left(m_{\cPqt,\text{cal}}\right)},
\]
where $\sigma\left(m_{\cPqt,\text{cal}}\right)$ is the statistical uncertainty on an individual $m_{\cPqt,\text{cal}}$ for a pseudo-experiment generated at $m_{\cPqt,\text{gen}}$.
As depicted in Fig.~\ref{fig:Pull}, we find a mass pull width of $1.005$ for the muon channel and $1.048$ for the electron channel.
Our method slightly underestimates the statistical
uncertainty of the measurement, and we incorporate the required corrections into
the evaluation of the final likelihoods.

\begin{figure}
\subfloat[]{\includegraphics[width=0.5\textwidth]{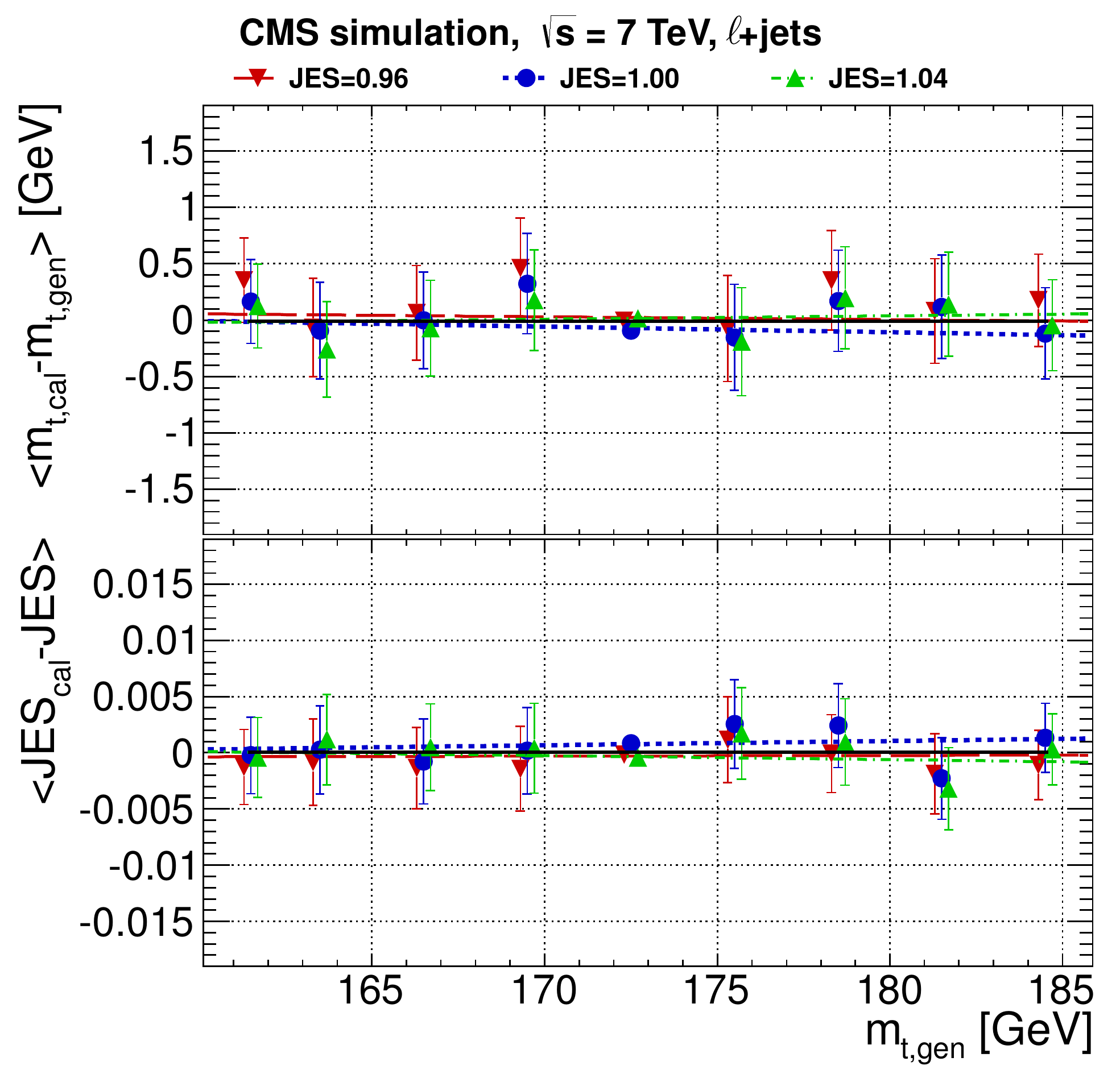}
}
\subfloat[]{\includegraphics[width=0.5\textwidth]{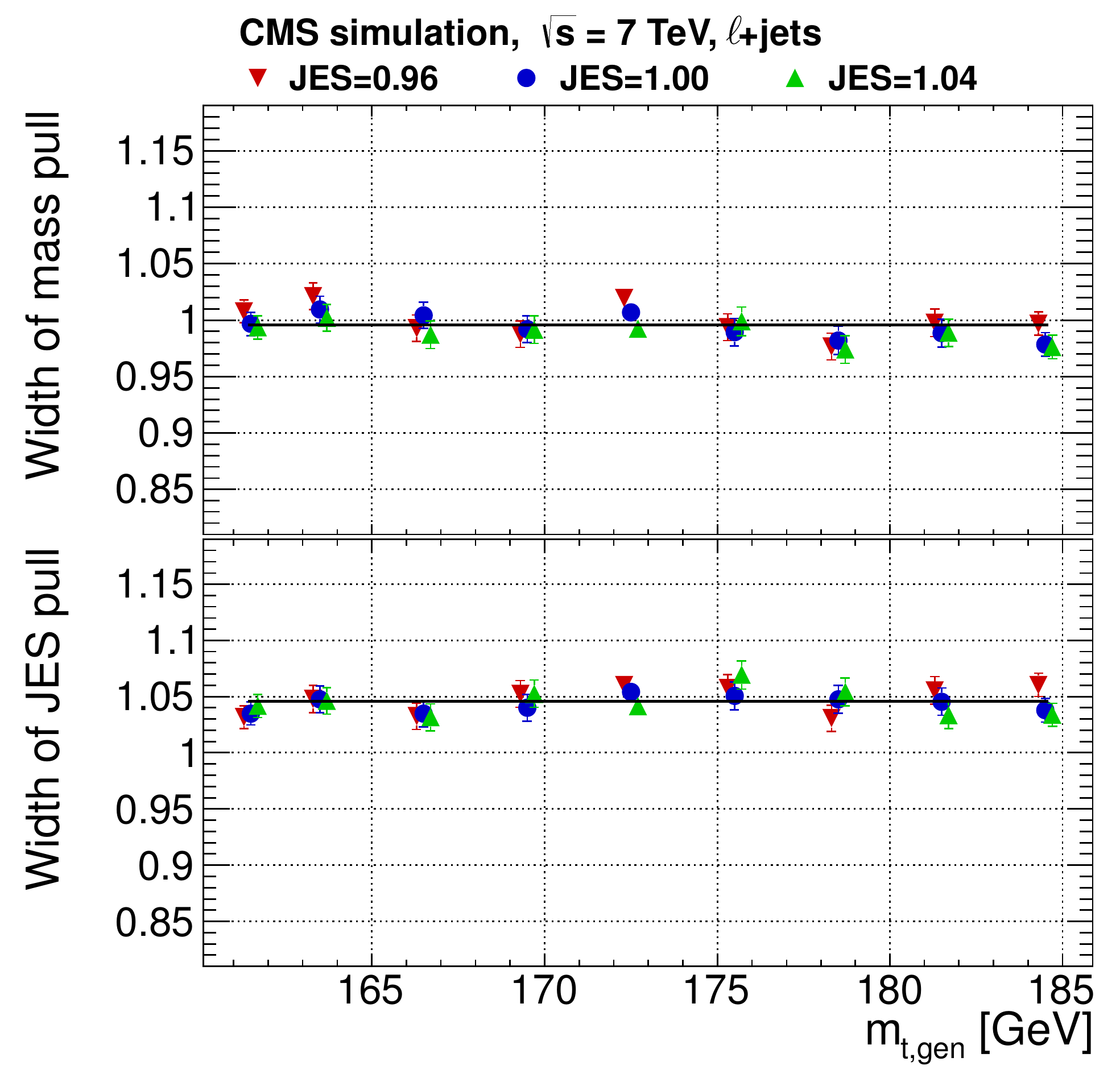}
}

\caption{\label{fig:Calibration_all} (a) Mean difference between the calibrated and generated values of $\mtop$ and JES as a function of different generated $m_{\cPqt,\text{gen}}$ and three values of JES for combined lepton+jets events; (b) width of the pull distributions for the combined channel after the single-channel calibration. The colored dashed lines correspond to straight line fits, the black solid line corresponds to a constant fit to all calibration points. The error bars in (a) indicate the statistical uncertainty on the mean difference for the uncalibrated likelihood.
}
\end{figure}

After applying the single-channel calibration, we again generate pseudo-experiments containing both muon and electron events to check the individual calibrations and the calibration for the combination of the two channels. As shown in Fig.~\ref{fig:Calibration_all}, statistical fluctuations are suppressed in the combination, and no additional corrections are needed.

\section{Systematic uncertainties}
\label{sec:systematics}
The contributions from the different sources of systematic uncertainties are shown in Table~\ref{tab:Systematic-uncertainties}, separately for the muon+jets and electron+jets final states, and for the combined fit to the entire data set.
In general, the absolute value of the largest observed shifts in $\mtop$ and JES, determined from changing the parameters by $\pm 1$ standard deviations, are assigned as systematic uncertainties on the final measurement.
The systematic uncertainties considered as relevant for this measurement, and the methods used to evaluate them are described below.

\begin{description}
\item [{{Fit calibration:}}] We propagate the statistical uncertainty on the calibration to the final measured quantities.
\item [{{b-JES:}}] The difference in the jet energy responses for jets originating from light (uds) or bottom quarks or gluons studied in simulation
indicate that the response to b-jets lies between the jet responses to light-quark and gluons~\cite{Chatrchyan:2011ds}.
Hence, the uncertainty assumed for the flavor dependence of the JES determination, which covers the transition from a gluon-dominated to a light quark-dominated sample, also covers the difference between light quarks and bottom quarks.
Thus, all the momenta of all b jets in simulation are scaled up and down by their individual flavor uncertainties.
\item [{{\pt- and $\eta$-dependent JES:}}] As we measure a constant jet
energy scale we have to take into account the influence of the \pt-
and $\eta$-dependent jet energy uncertainties. This is done by scaling the energies of all jets up and down according to their individual uncertainties~\cite{Chatrchyan:2011ds}. We take the largest difference in the measured top-quark mass and JES (compared to the expected average JES shift of 1.6\%) as a systematic uncertainty.
\item [{{Lepton energy scale:}}] We shift the muon and electron energies in simulation up and down according to their respective uncertainties.
\item [{{Missing transverse momentum:}}] In addition to propagating the jet and lepton energy scale uncertainties, we scale the unclustered energy up and down by $10$\%.
\item [{{Jet energy resolution:}}] The jet energy resolution in simulation
is degraded by $7$ to $20$\% depending on $\eta$ to match the resolutions measured in data in Ref.~\cite{Chatrchyan:2011ds}. To account for the resolution uncertainty, the jet energy resolution in the simulation is modified by ${\pm}1$ standard deviations with respect to the degraded resolution.
\item [{{b tagging:}}] The nominal requirement on the CSVM tagger is varied in order to reflect the uncertainty of the b-tag efficiency of $3$\%~\cite{CMS-PAS-BTV-11-004}.
\item [{{Pileup:}}] The effect of the pileup events is evaluated by weighting the simulation to provide as many additional minimum bias events as expected from the inelastic pp cross section. To cover the uncertainties associated with the number of pileup events, the average number of expected pileup events ($9.3$ for the analyzed data) is changed by ${\pm}5$\%.
\item [{{Non-\ttbar background:}}] After final selection, background fractions of $1$\% W+jets and $3$\% single top-quark events are expected from simulation.
We conduct pseudo-experiments with 2\% $\PW+\cPqb\cPaqb$ and 6\% single top-quark events and take the difference to the result without background as a systematic uncertainty.
\item [{{Parton distribution functions:}}] The simulated events are generated using the CTEQ 6.6L parton distribution functions (PDFs)~\cite{Nadolsky:2008zw}. The uncertainty on this set of PDFs is given in terms of variations of 22 orthogonal parameters that can be used to reweight the events, resulting in 22 pairs of additional PDFs.
Half of the difference in top-quark mass and JES of each pair is quoted as a systematic uncertainty.
\item [{{Renormalization and factorization scales:}}] The dependence of the result on the renormalization and factorization scales used in the \ttbar simulation is studied by changing the nominal renormalization and factorization scales simultaneously by factors of 0.5 and 2. This change in the parameters also reflects the uncertainty on the amount of initial- and final-state radiation.
\item [{{ME-PS matching threshold:}}] In the \ttbar simulation, the matching thresholds used for interfacing the matrix-elements (ME) generated with \MADGRAPH and the \PYTHIA parton showering (PS) are changed from the default value of 20\GeV down to 10\GeV and up to 40\GeV.
\item [{{Underlying event:}}] Non-pertubative QCD effects are taken into account by tuning \PYTHIA to measurements of the underlying event~\cite{Chatrchyan:2011id}. The uncertainties are estimated by comparing two tunes with increased and decreased underlying event activity relative to a central tune (the Perugia 2011 tune compared to the Perugia 2011 mpiHi and Perugia 2011 Tevatron tunes described in Ref.~\cite{Skands:2010ak}).
\item [{{Color reconnection effects:}}] The uncertainties that arise from ambiguities in modeling color reconnection effects~\cite{Skands:2007zg} are estimated by comparing in simulation an underlying event tune including color reconnection to a tune without it (the Perugia 2011 and Perugia 2011 noCR tunes described in Ref.~\cite{Skands:2010ak}).
\end{description}

\begin{table}
\topcaption{\label{tab:Systematic-uncertainties}List of systematic uncertainties for the muon+jets and electron+jets final states, and for the combined fit to the entire data set}
\noindent \centering{}%
\begin{tabular}{l|cc|cc|cc}
 & \multicolumn{2}{c|}{$\mu$+jets}
 & \multicolumn{2}{c|}{e+jets}
  & \multicolumn{2}{c}{$\ell$+jets}\tabularnewline
 & $\delta^{\mu}_{\mtop}$ (\GeVns) & $\delta^{\mu}_\mathrm{JES}$
 & $\delta^{\Pe}_{\mtop}$ (\GeVns) & $\delta^{\Pe}_\mathrm{JES}$
 & $\delta^{\ell}_{\mtop}$ (\GeVns) & $\delta^{\ell}_\mathrm{JES}$\tabularnewline
\hline
\hline
Fit calibration
 & 0.08 & 0.001 & 0.09 & 0.001 & 0.06 & 0.001\tabularnewline
b-JES
 & 0.60 & 0.000 & 0.62 & 0.000 & 0.61 & 0.000\tabularnewline
\pt- and $\eta$-dependent JES
 & 0.30 & 0.001 & 0.28 & 0.001 & 0.28 & 0.001\tabularnewline
Lepton energy scale
 & 0.03 & 0.000 & 0.04 & 0.000 & 0.02 & 0.000\tabularnewline
Missing transverse momentum
 & 0.05 & 0.000 & 0.07 & 0.000 & 0.06 & 0.000\tabularnewline
Jet energy resolution
 & 0.22 & 0.004 & 0.24 & 0.004 & 0.23 & 0.004\tabularnewline
b tagging
 & 0.11 & 0.001 & 0.15 & 0.001 & 0.12 & 0.001\tabularnewline
Pileup
 & 0.07 & 0.002 & 0.08 & 0.001 & 0.07 & 0.001\tabularnewline
Non-\ttbar background
 & 0.10 & 0.001 & 0.16 & 0.000 & 0.13 & 0.001\tabularnewline
Parton distribution functions
 & 0.07 & 0.001 & 0.07 & 0.001 & 0.07 & 0.001\tabularnewline
Renormalization and
 & \multirow{2}{*}{0.23} & \multirow{2}{*}{0.004} & \multirow{2}{*}{0.41} & \multirow{2}{*}{0.005} & \multirow{2}{*}{0.24} & \multirow{2}{*}{0.004}\tabularnewline
factorization scales & & & & & & \tabularnewline
ME-PS matching threshold
 & 0.17 & 0.000 & 0.15 & 0.001 & 0.18 & 0.001\tabularnewline
Underlying event
 & 0.26 & 0.002 & 0.24 & 0.001 & 0.15 & 0.002\tabularnewline
Color reconnection effects
 & 0.66 & 0.004 & 0.39 & 0.003 & 0.54 & 0.004\tabularnewline
\hline
\hline
Total & 1.06 & 0.008 & 1.00 & 0.007 & 0.98 & 0.008\tabularnewline
\end{tabular}
\end{table}

Differences in the systematic uncertainties calculated for the muon+jets and electron+jets final states are consistent with arising from statistical fluctuations due to the limited sizes of the studied samples.
The total systematic uncertainty is dominated by effects that cannot be compensated through a simultaneous determination of $\mtop$ and JES. Besides the JES for b quarks, the uncertainty in color reconnection dominates. For the sample without color reconnection effects, the mean of the $m_\PW^{\text{reco}}$ distribution shifts relative to the reference sample, leading to an additional uncertainty on $\mtop$ for the simultaneous fit.

\section{Measurement of the mass of the top quark}
\label{sec:results}
Using the selected samples, we measure:
\begin{align*}
\mbox{$\mu$+jets: } \mtop =&  173.22\pm0.56\,\text{(stat.+JES)}\pm1.06\,\text{(syst.)}\GeV,\  \mathrm{JES} = 0.999\pm0.005\,\text{(stat.)}\pm0.008\,\text{(syst.)},\\
\mbox{e+jets: } \mtop =&  173.72\pm0.66\,\text{(stat.+JES)}\pm1.00\,\text{(syst.)}\GeV,\  \mathrm{JES} = 0.989\pm0.005\,\text{(stat.)}\pm0.007\,\text{(syst.)}.
\end{align*}

The combined fit to the 5174 $\ell$+jets events in the two channels yields:
\begin{eqnarray*}
\mtop & = & 173.49\pm0.43\,\text{(stat.+JES)}\pm0.98\,\text{(syst.)}\GeV,\\
\mbox{JES} & = & 0.994\pm0.003\,\text{(stat.)}\pm0.008\,\text{(syst.)}.
\end{eqnarray*}

The overall uncertainty of the presented measurement is $1.07$\GeV on the top-quark mass from adding the components in quadrature. The measured JES confirms that obtained from events with Z bosons and photons~\cite{Chatrchyan:2011ds}.

\begin{figure}
\noindent \begin{centering}
\subfloat[\label{fig:result-data}]{\noindent \begin{centering}
\includegraphics[width=0.5\textwidth]{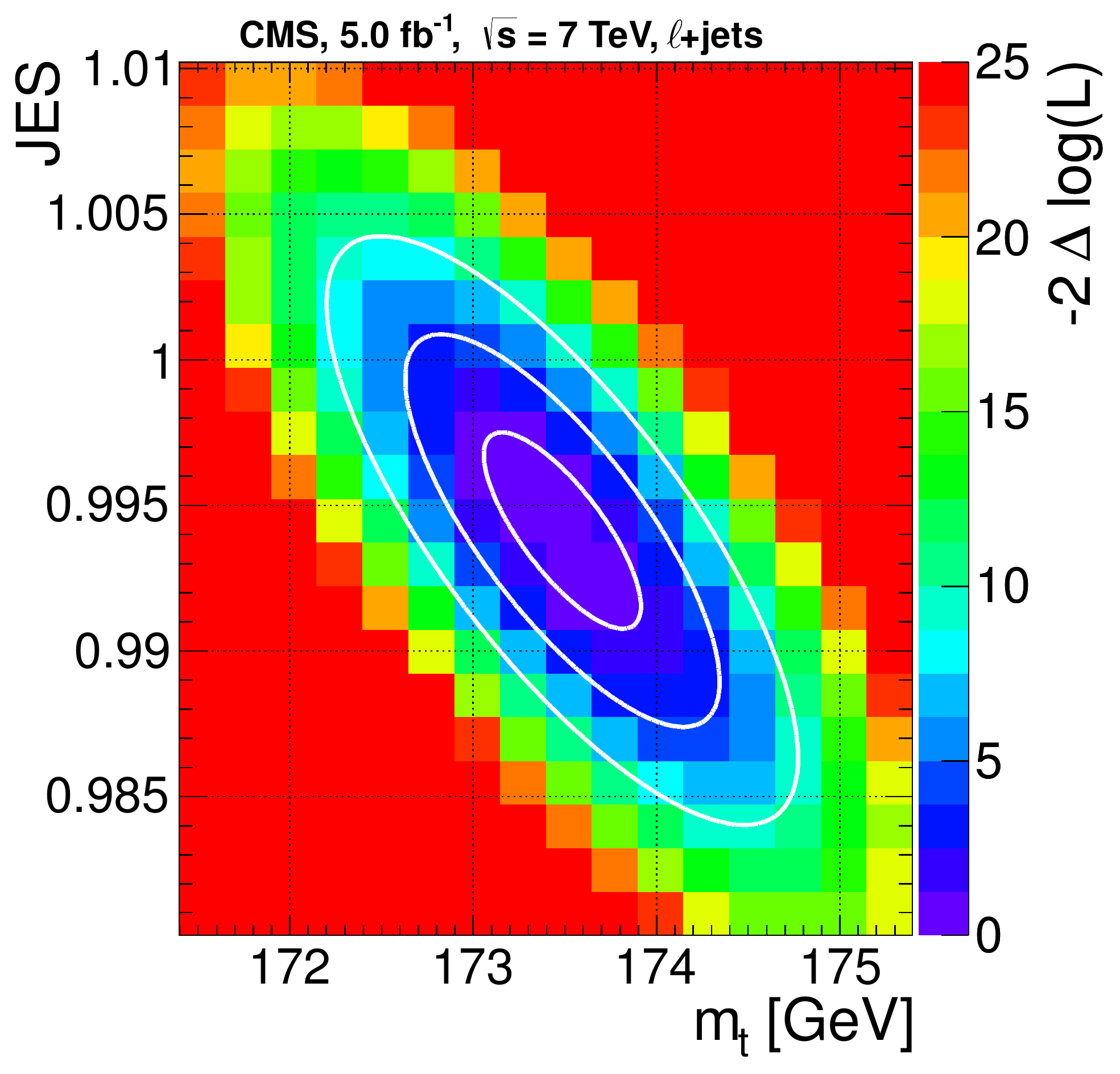}
\par\end{centering}

}\subfloat[\label{fig:result-unc}]{\noindent \begin{centering}
\includegraphics[width=0.5\textwidth]{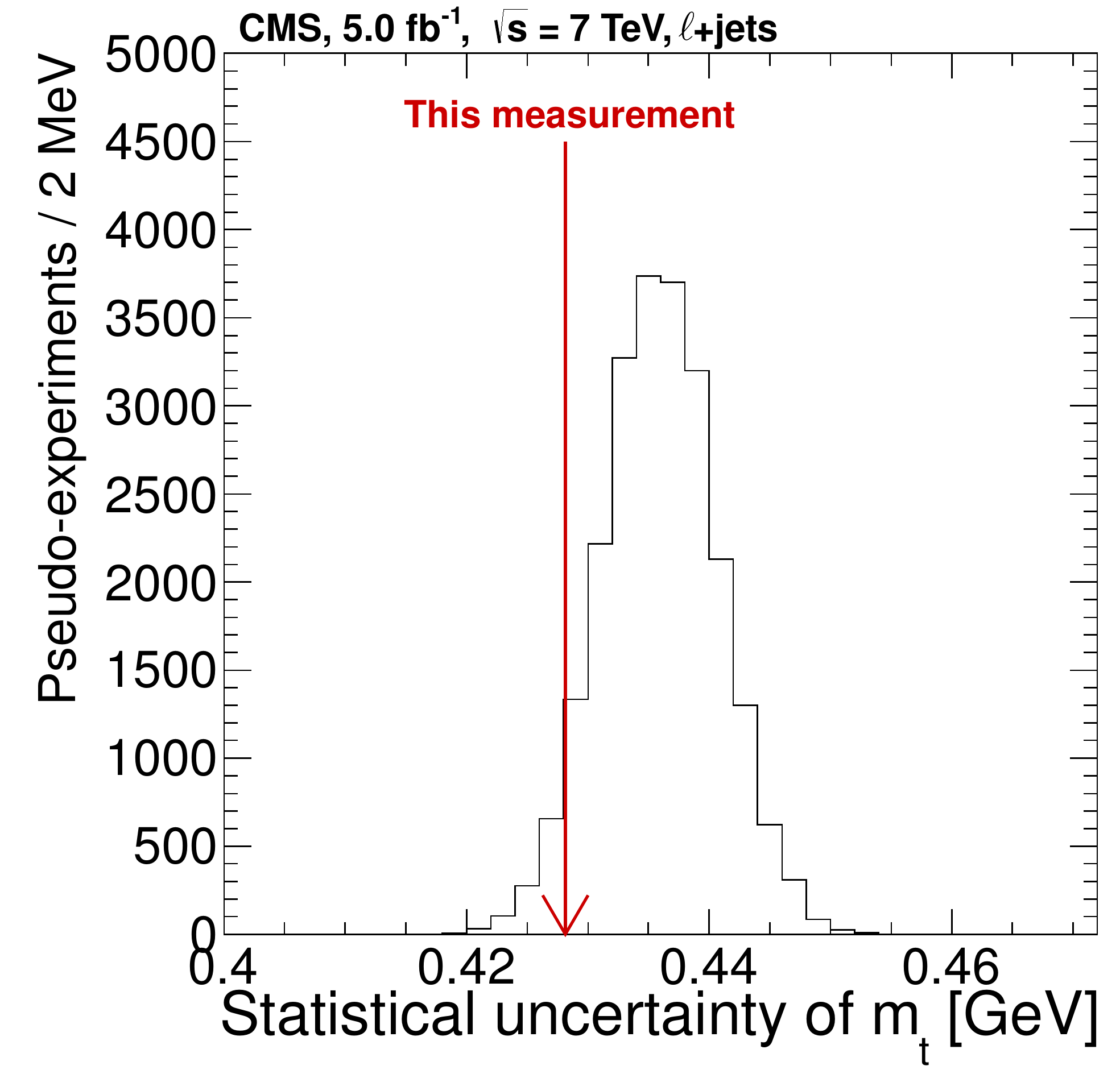}
\par\end{centering}

}
\par\end{centering}

\caption{\label{fig:result}(a) The 2D likelihood ($-2 \Delta \log \left(\mathcal{L}\right)$) measured for the $\ell$+jets final state.
The ellipses correspond to statistical uncertainties on $\mtop$ and JES of one, two, and three standard deviations. (b) The statistical uncertainty distribution obtained from 10\,000 pseudo-experiments is compared to the uncertainty of the measurement in data of $0.43\GeV$.}
\end{figure}

Figure~\ref{fig:result} (a) shows the 2D likelihood obtained from data. As depicted in Fig.~\ref{fig:result} (b), the uncertainty of the measurement agrees with the expected precision from the pseudo-experiments.
As the top-quark mass  and the JES are measured simultaneously, the statistical uncertainty on $\mtop$ combines the statistical uncertainty arising from both components of the measurement.

We estimate the impact of the simultaneous fit of the jet energy scale by fixing the JES to unity. This yields $ \mtop = 172.97 \pm 0.27\,\text{(stat.)} \pm 1.44\,\text{(syst.)\GeV.} $
The larger systematic uncertainty stems from a JES uncertainty of $1.33$\GeV and demonstrates the gain from the simultaneous fit to $\mtop$ and JES.

\begin{figure}[h]
\noindent \begin{centering}
\includegraphics[width=0.5\textwidth]{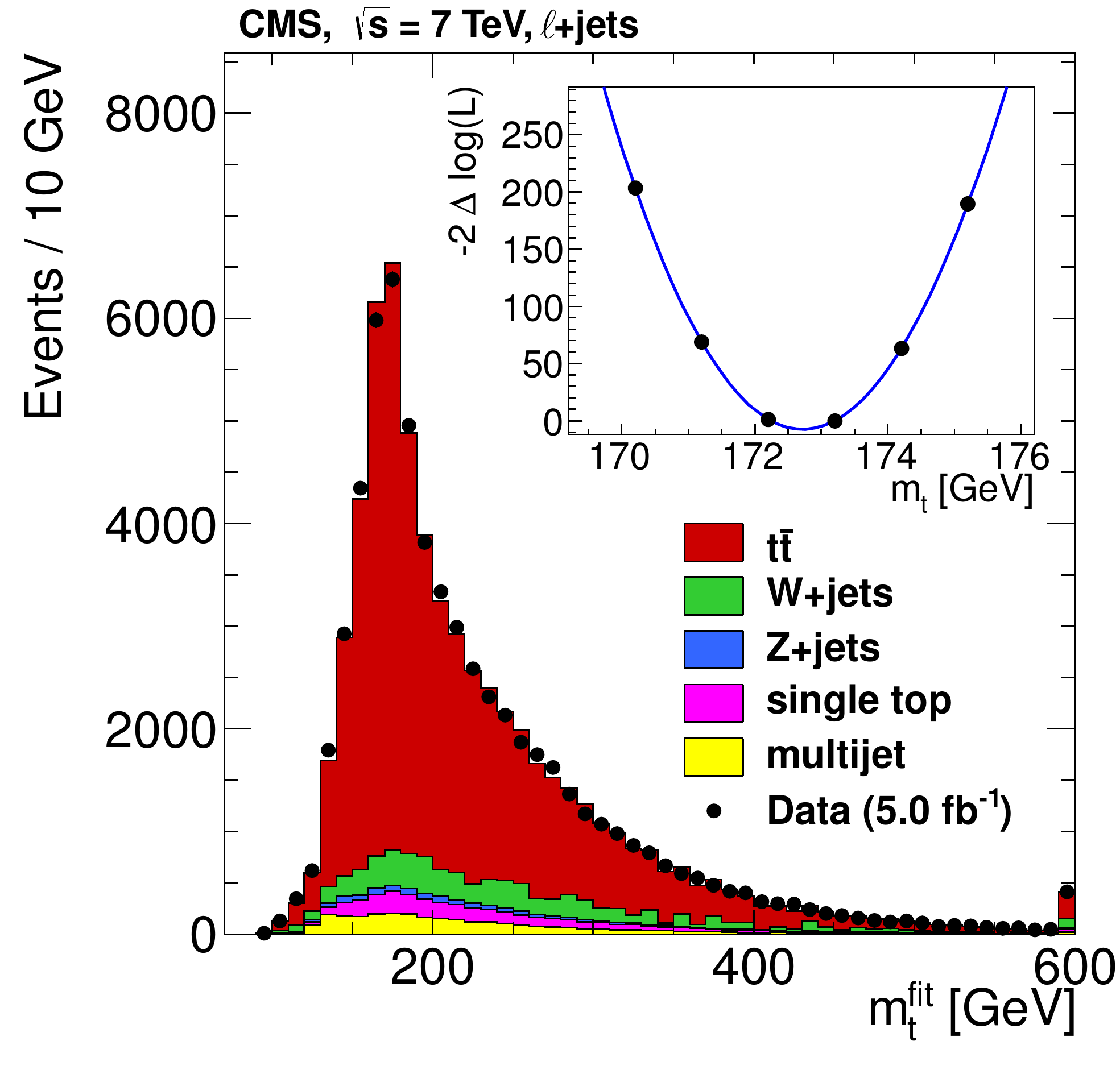}
\par\end{centering}

\caption{\label{fig:fittedMass} Distribution of the reconstructed top-quark mass after the kinematic fit, for the permutation with the lowest $\chi^2$, in the cross-check analysis. The distributions are normalized to the number of events observed in data. The top-quark mass assumed in the simulation is $172.5$\GeV.
The rightmost bin also contains the overflow. The inset shows the cross-check likelihood measured for the $\ell$+jets final state.
}
\end{figure}

As a cross-check of the event selection and the mass extraction technique, a second analysis is performed. The mass extraction technique used in this analysis is very similar to the CMS measurement of the mass difference between top and antitop quarks~\cite{Chatrchyan:2012ub}. The events are required to pass a lepton+jets trigger and fulfill the same lepton and jet requirements implemented in the main analysis, except for a lower threshold on the muon transverse momentum of $\pt>20$\GeV.
In addition, the requirement on the number of b-tagged jets is lowered to at least one.
The same kinematic fit is employed as for the main analysis. All possible permutations of the four jets of largest $\pt$ that have a fit $\chi^2 < 20$, corresponding to a goodness-of-fit probability of $P_\mathrm{gof} = 4.5\times 10^{-5}$, are accepted, yielding 54\,899 selected events.
 This fit to $\mtop$ employs just the standard jet energy corrections, equivalent to setting JES=1.
New likelihood functions are formed that take account of the contributions from background resulting from less stringent selection criteria. In addition, the permutations are weighted with the probabilities for b tagging~\cite{Chatrchyan:2012ub}.
Figure~\ref{fig:fittedMass} shows the mass $\mtop^\text{fit}$  after the kinematic fit for the permutation with the smallest $\chi^{2}$ in each event and the likelihood obtained from data.
After applying the calibration, we measure a top-quark mass of $ \mtop = 172.72 \pm 0.18\,\text{(stat.)} \pm 1.49\,\text{(syst.)\GeV}$, which is consistent with the result of the main analysis but has a larger uncertainty as expected.

We use the BLUE method~\cite{Lyons:1988rp} to combine the result presented in this letter with the measurements in the dilepton channel in 2010~\cite{Chatrchyan:2011nb} and 2011~\cite{CMS-TOP-11-016-001}.
Most of the systematic uncertainties listed in Table \ref{tab:Systematic-uncertainties} are assumed to be fully correlated among the three input measurements.
Exceptions are the uncertainties on pileup, for which we assign full correlation between the 2011 analyses but no correlation with the 2010 analysis, since the pileup conditions and their treatments differ.
In addition, the mass calibration, the statistical uncertainty on the \emph{in situ} fit for the JES, and the data-based background normalization in the dilepton analyses are treated as uncorrelated systematic uncertainties.
As a cross-check, the correlation coefficients for the correlated sources are varied simultaneously between 1 and 0, and changes smaller than $0.14\GeV$ are observed in the combined result.
The combination of the three measurements yields a mass of $\mtop = 173.32 \pm 0.27\,\text{(stat.)} \pm 1.02\,\text{(syst.)}$\GeV. It has a $\chi^2$ of 1.05 for two degrees of freedom, which corresponds to a probability of 59\%.

\section{Summary}
\label{sec:conclusion}
The complete kinematic properties of each event are reconstructed through a constrained fit to a $\ttbar$ hypothesis.
For each selected event, a likelihood is calculated as a function of assumed values of the top-quark mass and the jet energy scale, taking into account all possible jet assignments.
From a data sample corresponding to an integrated luminosity of $5.0$\fbinv, 5174 candidate events are selected and the top-quark mass is measured to be
$m_{\textrm{t}} = 173.49 \pm 0.43\,\text{(stat.+JES)} \pm 0.98\,\text{(syst.)\GeV.}$
This result is
consistent with the Tevatron average~\cite{Lancaster:2011wr},
and constitutes the most precise single measurement to date of the mass of the top quark.

\section*{Acknowledgements}
We congratulate our colleagues in the CERN accelerator departments for the excellent performance of the LHC machine. We thank the technical and administrative staff at CERN and other CMS institutes, and acknowledge support from BMWF and FWF (Austria); FNRS and FWO (Belgium); CNPq, CAPES, FAPERJ, and FAPESP (Brazil); MES (Bulgaria); CERN; CAS, MoST, and NSFC (China); COLCIENCIAS (Colombia); MSES (Croatia); RPF (Cyprus); MoER, SF0690030s09 and ERDF (Estonia); Academy of Finland, MEC, and HIP (Finland); CEA and CNRS/IN2P3 (France); BMBF, DFG, and HGF (Germany); GSRT (Greece); OTKA and NKTH (Hungary); DAE and DST (India); IPM (Iran); SFI (Ireland); INFN (Italy); NRF and WCU (Korea); LAS (Lithuania); CINVESTAV, CONACYT, SEP, and UASLP-FAI (Mexico); MSI (New Zealand); PAEC (Pakistan); MSHE and NSC (Poland); FCT (Portugal); JINR (Armenia, Belarus, Georgia, Ukraine, Uzbekistan); MON, RosAtom, RAS and RFBR (Russia); MSTD (Serbia); SEIDI and CPAN (Spain); Swiss Funding Agencies (Switzerland); NSC (Taipei); ThEP, IPST and NECTEC (Thailand); TUBITAK and TAEK (Turkey); NASU (Ukraine); STFC (United Kingdom); DOE and NSF (USA).

\bibliography{auto_generated}   

\cleardoublepage \appendix\section{The CMS Collaboration \label{app:collab}}\begin{sloppypar}\hyphenpenalty=5000\widowpenalty=500\clubpenalty=5000\input{TOP-11-015-authorlist.tex}\end{sloppypar}
\end{document}

%% file: TOP-11-015-authorlist.tex
\textbf{Yerevan Physics Institute,  Yerevan,  Armenia}\\*[0pt]
S.~Chatrchyan, V.~Khachatryan, A.M.~Sirunyan, A.~Tumasyan
\vskip\cmsinstskip
\textbf{Institut f\"{u}r Hochenergiephysik der OeAW,  Wien,  Austria}\\*[0pt]
W.~Adam, E.~Aguilo, T.~Bergauer, M.~Dragicevic, J.~Er\"{o}, C.~Fabjan\cmsAuthorMark{1}, M.~Friedl, R.~Fr\"{u}hwirth\cmsAuthorMark{1}, V.M.~Ghete, J.~Hammer, N.~H\"{o}rmann, J.~Hrubec, M.~Jeitler\cmsAuthorMark{1}, W.~Kiesenhofer, V.~Kn\"{u}nz, M.~Krammer\cmsAuthorMark{1}, I.~Kr\"{a}tschmer, D.~Liko, I.~Mikulec, M.~Pernicka$^{\textrm{\dag}}$, B.~Rahbaran, C.~Rohringer, H.~Rohringer, R.~Sch\"{o}fbeck, J.~Strauss, A.~Taurok, W.~Waltenberger, G.~Walzel, E.~Widl, C.-E.~Wulz\cmsAuthorMark{1}
% ...
\vskip\cmsinstskip
\dag:~Deceased\\
1:~~Also at Vienna University of Technology, Vienna, Austria\\
% ...


%% file: TOP-11-015_temp.bbl
\providecommand{\href}[2]{#2}\begingroup\raggedright\begin{thebibliography}{10}%
\makeatletter
\providecommand{\hrefCMSnoop }[0]{\@secondoftwo}%
\makeatother
\providecommand{\doi}{\texttt{doi:}\begingroup \urlstyle{tt}\Url}

\bibitem{CMS:2008zzk}
\hrefCMSnoop {} {{ {CMS}} Collaboration, ``The CMS experiment at the CERN
  LHC'',} \textit{ JINST} \textbf{ 3} (2008) S08004,
\href{http://dx.doi.org/10.1088/1748-0221/3/08/S08004}{\doi{10.1088/1748-0221/3/08/S08004}}.

\bibitem{Lancaster:2011wr}
\hrefCMSnoop {} {{CDF and \DZERO Collaborations}, ``{Combination of the
  top-quark mass measurements from the Tevatron collider}'',} (2012).
  \href{http://www.arXiv.org/abs/1207.1069}{\texttt{ arXiv:1207.1069}}.
Submitted to \textit{Phys. Rev. D}.

\bibitem{Aaltonen:2012zzz}
\hrefCMSnoop {} {{ CDF} Collaboration, ``{Precision Top-Quark Mass Measurement
  at CDF}'',} \textit{ Phys. Rev. Lett.} \textbf{ 109} (2012) 152003,
  \href{http://dx.doi.org/10.1103/PhysRevLett.109.152003}{\doi{10.1103/PhysRevLett.109.152003}},
\href{http://www.arXiv.org/abs/1207.6758}{\texttt{ arXiv:1207.6758}}.

\bibitem{Chatrchyan:2011nb}
\hrefCMSnoop {} {{ CMS} Collaboration, ``{Measurement of the \ttbar production
  cross section and the top quark mass in the dilepton channel in pp collisions
  at $\sqrt{s} = 7$~\TeV}'',} \textit{ JHEP} \textbf{ 07} (2011) 049,
  \href{http://dx.doi.org/10.1007/JHEP07(2011)049}{\doi{10.1007/JHEP07(2011)049}},
\href{http://www.arXiv.org/abs/1105.5661}{\texttt{ arXiv:1105.5661}}.

\bibitem{CMS-TOP-11-016-001}
\hrefCMSnoop {} {{ CMS} Collaboration, ``Measurement of the top-quark mass in
  \ttbar events with dilepton final state in pp collisions at $\sqrt{s}$ = 7
  TeV'',} (2012). \href{http://www.arXiv.org/abs/1209.2393}{\texttt{
  arXiv:1209.2393}}. Submitted to \textit{Eur. Phys. J. C}.

\bibitem{Aad:2012aa}
\hrefCMSnoop {} {{ ATLAS} Collaboration, ``{Measurement of the top quark mass
  with the template method in the \ttbar $\to$ lepton + jets channel using
  ATLAS data}'',} \textit{ Eur. Phys. J. C} \textbf{ 72} (2012) 2046,
  \href{http://dx.doi.org/10.1140/epjc/s10052-012-2046-6}{\doi{10.1140/epjc/s10052-012-2046-6}},
\href{http://www.arXiv.org/abs/1203.5755}{\texttt{ arXiv:1203.5755}}.

\bibitem{pdg}
\hrefCMSnoop {} {{ {Particle Data Group}} Collaboration, ``{Review of Particle
  Physics}'',} \textit{ Phys. Rev. D} \textbf{ 86} (2012) 010001,
  \href{http://dx.doi.org/10.1103/PhysRevD.86.010001}{\doi{10.1103/PhysRevD.86.010001}}.

\bibitem{Alwall:2011uj}
J.~Alwall\hrefCMSnoop {} { {et~al.}, ``{MadGraph 5 : going beyond}'',} \textit{
  JHEP} \textbf{ 06} (2011) 128,
  \href{http://dx.doi.org/10.1007/JHEP06(2011)128}{\doi{10.1007/JHEP06(2011)128}},
\href{http://www.arXiv.org/abs/1106.0522}{\texttt{ arXiv:1106.0522}}.

\bibitem{Sjostrand:2006za}
\hrefCMSnoop {} {T.~{Sj\"ostrand}, S.~Mrenna, and P.~Z. Skands, ``{\PYTHIA 6.4
  physics and manual}'',} \textit{ JHEP} \textbf{ 05} (2006) 026,
  \href{http://dx.doi.org/10.1088/1126-6708/2006/05/026}{\doi{10.1088/1126-6708/2006/05/026}},
\href{http://www.arXiv.org/abs/hep-ph/0603175}{\texttt{ arXiv:hep-ph/0603175}}.

\bibitem{Chatrchyan:2011id}
\hrefCMSnoop {} {{ CMS} Collaboration, ``{Measurement of the underlying event
  activity at the LHC with $\sqrt{s}= 7$ TeV and comparison with $\sqrt{s} =
  0.9$ TeV}'',} \textit{ JHEP} \textbf{ 09} (2011) 109,
  \href{http://dx.doi.org/10.1007/JHEP09(2011)109}{\doi{10.1007/JHEP09(2011)109}},
\href{http://www.arXiv.org/abs/1107.0330}{\texttt{ arXiv:1107.0330}}.

\bibitem{Agostinelli:2002hh}
\hrefCMSnoop {} {{ GEANT4} Collaboration, ``{\GEANT4 -- a simulation
  toolkit}'',} \textit{ Nucl. Instrum. Meth. A} \textbf{ 506} (2003) 250,
\href{http://dx.doi.org/10.1016/S0168-9002(03)01368-8}{\doi{10.1016/S0168-9002(03)01368-8}}.

\bibitem{Alioli:2009je}
S.~Alioli\hrefCMSnoop {} { {et~al.}, ``{NLO single-top production matched with
  shower in \POWHEG: $s$- and $t$-channel contributions}'',} \textit{ JHEP}
  \textbf{ 09} (2009) 111,
  \href{http://dx.doi.org/10.1088/1126-6708/2009/09/111}{\doi{10.1088/1126-6708/2009/09/111}},
  \href{http://www.arXiv.org/abs/0907.4076}{\texttt{ arXiv:0907.4076}}.

\bibitem{Re:2010bp}
\hrefCMSnoop {} {E.~Re, ``{Single-top Wt-channel production matched with parton
  showers using the \POWHEG method}'',} \textit{ Eur. Phys. J. C} \textbf{ 71}
  (2011) 1547,
  \href{http://dx.doi.org/10.1140/epjc/s10052-011-1547-z}{\doi{10.1140/epjc/s10052-011-1547-z}},
  \href{http://www.arXiv.org/abs/1009.2450}{\texttt{ arXiv:1009.2450}}.

\bibitem{Nason:2004rx}
\hrefCMSnoop {} {P.~Nason, ``{A new method for combining NLO QCD with shower
  Monte Carlo algorithms}'',} \textit{ JHEP} \textbf{ 11} (2004) 040,
  \href{http://dx.doi.org/10.1088/1126-6708/2004/11/040}{\doi{10.1088/1126-6708/2004/11/040}},
  \href{http://www.arXiv.org/abs/hep-ph/0409146}{\texttt{
  arXiv:hep-ph/0409146}}.

\bibitem{Frixione:2007vw}
\hrefCMSnoop {} {S.~Frixione, P.~Nason, and C.~Oleari, ``{Matching NLO QCD
  computations with parton shower simulations: the \POWHEG method}'',} \textit{
  JHEP} \textbf{ 11} (2007) 070,
  \href{http://dx.doi.org/10.1088/1126-6708/2007/11/070}{\doi{10.1088/1126-6708/2007/11/070}},
  \href{http://www.arXiv.org/abs/0709.2092}{\texttt{ arXiv:0709.2092}}.

\bibitem{Alioli:2010xd}
S.~Alioli\hrefCMSnoop {} { {et~al.}, ``{A general framework for implementing
  NLO calculations in shower Monte Carlo programs: the \POWHEG BOX}'',}
  \textit{ JHEP} \textbf{ 06} (2010) 043,
  \href{http://dx.doi.org/10.1007/JHEP06(2010)043}{\doi{10.1007/JHEP06(2010)043}},
  \href{http://www.arXiv.org/abs/1002.2581}{\texttt{ arXiv:1002.2581}}.

\bibitem{Kidonakis:2010dk}
\hrefCMSnoop {} {N.~Kidonakis, ``{Next-to-next-to-leading soft-gluon
  corrections for the top quark cross section and transverse momentum
  distribution}'',} \textit{ Phys. Rev. D} \textbf{ 82} (2010) 114030,
  \href{http://dx.doi.org/10.1103/PhysRevD.82.114030}{\doi{10.1103/PhysRevD.82.114030}},
  \href{http://www.arXiv.org/abs/1009.4935}{\texttt{ arXiv:1009.4935}}.

\bibitem{Melnikov:2006kv}
\hrefCMSnoop {} {K.~Melnikov and F.~Petriello, ``{Electroweak gauge boson
  production at hadron colliders through $O(\alpha_s^2)$}'',} \textit{ Phys.
  Rev. D} \textbf{ 74} (2006) 114017,
  \href{http://dx.doi.org/10.1103/PhysRevD.74.114017}{\doi{10.1103/PhysRevD.74.114017}},
\href{http://www.arXiv.org/abs/hep-ph/0609070}{\texttt{ arXiv:hep-ph/0609070}}.

\bibitem{Campbell:2010ff}
\hrefCMSnoop {} {J.~M. Campbell and R.~K. Ellis, ``{MCFM for the Tevatron and
  the LHC}'',} \textit{ Nucl. Phys. Proc. Suppl.} \textbf{ 205-206} (2010) 10,
  \href{http://dx.doi.org/10.1016/j.nuclphysbps.2010.08.011}{\doi{10.1016/j.nuclphysbps.2010.08.011}},
  \href{http://www.arXiv.org/abs/1007.3492}{\texttt{ arXiv:1007.3492}}.

\bibitem{Chatrchyan:2011ds}
\hrefCMSnoop {} {{ CMS} Collaboration, ``{Determination of jet energy
  calibration and transverse momentum resolution in CMS}'',} \textit{ JINST}
  \textbf{ 6} (2011) P11002,
  \href{http://dx.doi.org/10.1088/1748-0221/6/11/P11002}{\doi{10.1088/1748-0221/6/11/P11002}},
\href{http://www.arXiv.org/abs/1107.4277}{\texttt{ arXiv:1107.4277}}.

\bibitem{CMS-PAS-PFT-10-002}
\href {http://cdsweb.cern.ch/record/1279341} {{ CMS} Collaboration,
  ``Commissioning of the Particle-Flow Reconstruction in Minimum-Bias and Jet
  Events from {\Pp\Pp} Collisions at 7 {TeV}'',} CMS Physics Analysis Summary
  CMS-PAS-PFT-10-002, CERN, (2010).

\bibitem{Cacciari:2005hq}
\hrefCMSnoop {} {M.~Cacciari and G.~P. Salam, ``{Dispelling the $N^{3}$ myth
  for the \kt jet-finder}'',} \textit{ Phys. Lett. B} \textbf{ 641} (2006) 57,
  \href{http://dx.doi.org/10.1016/j.physletb.2006.08.037}{\doi{10.1016/j.physletb.2006.08.037}},
\href{http://www.arXiv.org/abs/hep-ph/0512210}{\texttt{ arXiv:hep-ph/0512210}}.

\bibitem{Cacciari:2008gp}
\hrefCMSnoop {} {M.~Cacciari, G.~P. Salam, and G.~Soyez, ``{The anti-\kt jet
  clustering algorithm}'',} \textit{ JHEP} \textbf{ 04} (2008) 063,
  \href{http://dx.doi.org/10.1088/1126-6708/2008/04/063}{\doi{10.1088/1126-6708/2008/04/063}},
\href{http://www.arXiv.org/abs/0802.1189}{\texttt{ arXiv:0802.1189}}.

\bibitem{Cacciari:2007fd}
\hrefCMSnoop {} {M.~Cacciari and G.~P. Salam, ``{Pileup subtraction using jet
  areas}'',} \textit{ Phys. Lett. B} \textbf{ 659} (2008) 119,
\href{http://dx.doi.org/10.1016/j.physletb.2007.09.077}{\doi{10.1016/j.physletb.2007.09.077}}.

\bibitem{Cacciari:2008gn}
\hrefCMSnoop {} {M.~Cacciari, G.~P. Salam, and G.~Soyez, ``{The catchment area
  of jets}'',} \textit{ JHEP} \textbf{ 04} (2008) 005,
\href{http://dx.doi.org/10.1088/1126-6708/2008/04/005}{\doi{10.1088/1126-6708/2008/04/005}}.

\bibitem{Cacciari:2011ma}
\hrefCMSnoop {} {M.~Cacciari, G.~P. Salam, and G.~Soyez, ``{FastJet user
  manual}'',} (2011). \href{http://www.arXiv.org/abs/1111.6097}{\texttt{
  arXiv:1111.6097}}.
{}.

\bibitem{CMS-PAS-BTV-11-004}
\href {http://cdsweb.cern.ch/record/1427247} {{ CMS} Collaboration, ``b-Jet
  Identification in the {CMS} Experiment'',} CMS Physics Analysis Summary
  CMS-PAS-BTV-11-004, CERN, (2011).

\bibitem{Chatrchyan:2011yy}
\hrefCMSnoop {} {{ CMS} Collaboration, ``{Measurement of the \ttbar production
  cross section in pp Collisions at 7 TeV in lepton + jets events using b-quark
  jet identification}'',} \textit{ Phys. Rev. D} \textbf{ 84} (2011) 092004,
  \href{http://dx.doi.org/10.1103/PhysRevD.84.092004}{\doi{10.1103/PhysRevD.84.092004}},
\href{http://www.arXiv.org/abs/1108.3773}{\texttt{ arXiv:1108.3773}}.

\bibitem{Chatrchyan:2012ub}
\hrefCMSnoop {} {{ CMS} Collaboration, ``{Measurement of the mass difference
  between top and antitop quarks}'',} \textit{ JHEP} \textbf{ 06} (2012) 109,
  \href{http://dx.doi.org/10.1007/JHEP06(2012)109}{\doi{10.1007/JHEP06(2012)109}},
\href{http://www.arXiv.org/abs/1204.2807}{\texttt{ arXiv:1204.2807}}.

\bibitem{Abbott:1998dc}
\hrefCMSnoop {} {{ \DZERO} Collaboration, ``{Direct measurement of the top
  quark mass at \DZERO}'',} \textit{ Phys. Rev. D} \textbf{ 58} (1998) 052001,
  \href{http://dx.doi.org/10.1103/PhysRevD.58.052001}{\doi{10.1103/PhysRevD.58.052001}},
\href{http://www.arXiv.org/abs/hep-ex/9801025}{\texttt{ arXiv:hep-ex/9801025}}.

\bibitem{Abazov:2007rk}
\hrefCMSnoop {} {{ \DZERO} Collaboration, ``{Measurement of the top quark mass
  in the lepton + jets channel using the ideogram method}'',} \textit{ Phys.
  Rev. D} \textbf{ 75} (2007) 092001,
  \href{http://dx.doi.org/10.1103/PhysRevD.75.092001}{\doi{10.1103/PhysRevD.75.092001}},
\href{http://www.arXiv.org/abs/hep-ex/0702018}{\texttt{ arXiv:hep-ex/0702018}}.

\bibitem{Abdallah:2008xh}
\hrefCMSnoop {} {{ DELPHI} Collaboration, ``{Measurement of the mass and width
  of the W boson in ${\rm e}^{+}{\rm e}^{-}$ collisions at $\sqrt{s}$ = 161 --
  209 GeV}'',} \textit{ Eur. Phys. J. C} \textbf{ 55} (2008) 1,
  \href{http://dx.doi.org/10.1140/epjc/s10052-008-0585-7}{\doi{10.1140/epjc/s10052-008-0585-7}},
\href{http://www.arXiv.org/abs/0803.2534}{\texttt{ arXiv:0803.2534}}.

\bibitem{Aaltonen:2006xc}
\hrefCMSnoop {} {{ CDF} Collaboration, ``{Measurement of the top-quark mass in
  all-hadronic decays in p$\bar{\rm p}$ collisions at CDF II}'',} \textit{
  Phys. Rev. Lett.} \textbf{ 98} (2007) 142001,
  \href{http://dx.doi.org/10.1103/PhysRevLett.98.142001}{\doi{10.1103/PhysRevLett.98.142001}},
\href{http://www.arXiv.org/abs/hep-ex/0612026}{\texttt{ arXiv:hep-ex/0612026}}.

\bibitem{Nadolsky:2008zw}
P.~M. Nadolsky\hrefCMSnoop {} { {et~al.}, ``{Implications of CTEQ global
  analysis for collider observables}'',} \textit{ Phys. Rev. D} \textbf{ 78}
  (2008) 013004,
  \href{http://dx.doi.org/10.1103/PhysRevD.78.013004}{\doi{10.1103/PhysRevD.78.013004}},
\href{http://www.arXiv.org/abs/0802.0007}{\texttt{ arXiv:0802.0007}}.

\bibitem{Skands:2010ak}
\hrefCMSnoop {} {P.~Z. Skands, ``{Tuning Monte Carlo generators: The Perugia
  tunes}'',} \textit{ Phys. Rev. D} \textbf{ 82} (2010) 074018,
  \href{http://dx.doi.org/10.1103/PhysRevD.82.074018}{\doi{10.1103/PhysRevD.82.074018}},
\href{http://www.arXiv.org/abs/1005.3457}{\texttt{ arXiv:1005.3457}}.

\bibitem{Skands:2007zg}
\hrefCMSnoop {} {P.~Z. Skands and D.~Wicke, ``{Non-perturbative QCD effects and
  the top mass at the Tevatron}'',} \textit{ Eur. Phys. J. C} \textbf{ 52}
  (2007) 133,
  \href{http://dx.doi.org/10.1140/epjc/s10052-007-0352-1}{\doi{10.1140/epjc/s10052-007-0352-1}},
\href{http://www.arXiv.org/abs/hep-ph/0703081}{\texttt{ arXiv:hep-ph/0703081}}.

\bibitem{Lyons:1988rp}
\hrefCMSnoop {} {L.~Lyons, D.~Gibaut, and P.~Clifford, ``{How to combine
  correlated estimates of a single physical quantity}'',} \textit{ Nucl.
  Instrum. Meth. A} \textbf{ 270} (1988) 110,
\href{http://dx.doi.org/10.1016/0168-9002(88)90018-6}{\doi{10.1016/0168-9002(88)90018-6}}.

\end{thebibliography}\endgroup
